\documentclass[preprint,12pt]{elsarticle}
\usepackage{graphicx}
\usepackage{epstopdf, epsfig}
\usepackage{amsmath}
\usepackage{amssymb}

\journal{European Journal of Mechancis. B/Fluids}

\begin{document}

\begin{frontmatter}

\title{Nonlinear water waves in shallow water in the presence of constant vorticity: A Whitham approach}

\author{Christian Kharif and Malek Abid} 

\address{
Aix Marseille Universit\'e, CNRS, Centrale Marseille\\
IRPHE UMR 7342\\
F-13384, Marseille, France
}

\begin{abstract}
Two-dimensional nonlinear gravity waves travelling in shallow water on a vertically sheared current of constant vorticity are considered. Using Euler equations, in the shallow water approximation, hyperbolic equations for the surface elevation and the horizontal velocity are derived. Using Riemann invariants of these equations, that are obtained analytically, a closed-form nonlinear evolution equation for the surface elevation is derived. A dispersive term is added to this equation using the exact linear dispersion relation. With this new single first-order partial differential equation, vorticity effects on undular bores are studied. Within the framework of weakly nonlinear waves, a KdV-type equation and a Whitham equation with constant vorticity are derived from this new model and the effect of vorticity on solitary waves and periodic waves is considered. Futhermore, within the framework of the new model and the Whitham equation a study of the effect of vorticity on the breaking time of dispersive waves and hyperbolic waves as well is carried out.
\end{abstract}


\end{frontmatter}

\section{Introduction}
Generally, in coastal and ocean waters, current velocity profiles are established by bottom friction and wind stress at the sea surface, and consequently are vertically varying. Ebb and flood currents due to the tide may have an important effect on water wave properties. In any region where the wind blows, the generated current affects the behavior  of the waves. The present work focuses on the nonlinear evolution of two-dimensional gravity waves propagating in shallow water on a shear current which varies linearly with depth. Consequently, the waves are travelling on a flow of constant vorticity. Considering constant vorticity is an approximation which allows the analytical derivation of a unique partial differential equation governing the nonlinear evolution of hyperbolic or dispersive waves in shallow water.
\newline 
Within the framework of long waves, there are very few papers devoted to unsteady nonlinear evolution of water waves propagating on an underlying vertically sheared current. \citet{Freeman} derived a KdV equation governing the time evolution of gravity waves on shear flows of arbitrary vorticity distribution. Using an asymptotic expansion method \citet{Choi} derived a Green-Naghdi system for long gravity waves in uniform shear flows (constant vorticity) and for weakly nonlinear waves he deduced from this system a Boussinesq-type equation and a KdV equation. The derivation of a Boussinesq-type equation and a Camassa-Holm equation with constant vorticity was carried out by \citet{Johnson}. Very recently, \citet{Richard} derived a dispersive shallow water model which is a generalisation of the classical Green-Naghdi model to the case of shear flows. \citet{Castro} investigated rigorously a Green-Naghdi type system including a general vorticity. At the same time and separately \citet{Kharif} and \citet{Hur} have derived shallow water wave equations with constant vorticity. \citet{Kharif} have considered the effect of a vertically sheared current on rogue wave properties whereas \citet{Hur} investigated the effect of vorticity on modulational instability and onset of breaking.
\newline
In the coastal zone where the vorticity is an important ingredient, water wave dynamics is governed by the Euler equations with boundary conditions at the free surface which is nonlinear and unknown {\em a priori}. Numerical integration of this system of equations is not a trivial task and so more simple models have been derived in the past to describe and investigate the dynamics of water waves phenomena in shallow water such as undular bores and nonlinear long wave propagation. For a review one can refer to the paper by \citet{Lannes}.
\newline
Following \citet{Whitham}, we propose a new model derived from the Euler equation for water waves propagating on a vertically sheared current of constant vorticity in shallow water. This new approach is easier to handle numerically. The heuristic introduction of dispersion allows the study of strongly nonlinear two-dimensional long gravity waves in the presence of vorticity and as well that of undular bores. From this model we derive, within the framework of weakly nonlinear waves satisfying the exact linear dispersion a Whitham equation and for weakly nonlinear and weakly dispersive waves the KdV equation previously obtained by \citet{Freeman} and \citet{Choi}. These different equations are then used to investigate the effect of constant vorticity on breaking time of dispersive waves and hyperbolic waves as well.

\section{Derivation of the new approach: The generalised Whitham equation}
We consider two-dimensional gravity water waves propagating at the free surface of a vertically sheared current of uniform intensity $\Omega$ which is the opposite of the vorticity. The wave train moves along the $x-\mathrm{axis}$ and the $z-\mathrm{axis}$ is oriented upward opposite to the gravity. The origin $z=0$ is the undisturbed free surface and $z=-h(x)$ is the rigid bottom. 
\vspace{0.2cm}
\newline
The continuity equation is
\begin{equation}
u_x+w_z=0 
\label{continuity}
\end{equation}
where $u$ and $w$ are the longitudinal and vertical components of the wave induced velocity, 
respectively. The underlying current is $U=U_0+\Omega z$ where $U_0$ is the constant surface velocity. 
\vspace{0.2cm}
\newline
Integration of equation (\ref{continuity}) gives
\begin{equation}
w(z=\eta)-w(z=-h)=-\int_{-h(x)}^{\eta(x,t)} u_x dz 
\label{interal-continuite}
\end{equation}
where $\eta$ is the surface elevation.
\vspace{0.2cm}
\newline
Note that
\begin{equation}
\int_{-h(x)}^{\eta(x,t)} u_x dz=\frac{\partial}{\partial x}\int_{-h(x)}^{\eta(x,t)} u dz-u(z=\eta)\eta_x-u(z=-h)h_x
\label{interal-continuite-2}
\end{equation}
\vspace{0.2cm}
\newline
The kinematic boundary condition at the free surface is
\begin{equation}
\eta_t+(u+U_0+\Omega \eta)\eta_x-w=0 \qquad \mathrm{on} \qquad z=\eta(x,t)
\label{kinematic}
\end{equation}
\vspace{0.2cm}
\newline
The bottom boundary condition writes
\begin{equation}
(u+U_0-\Omega h)h_x+w=0 \qquad \mathrm{on} \qquad z=-h(x)
\end{equation}
\vspace{0.2cm}
\newline
We assume $h$ constant, then
\begin{equation}
w=0 \qquad \qquad \mathrm{on} \qquad \qquad z=-h
\end{equation} 
\vspace{0.3cm}
\newline
From equation (\ref{kinematic}) it follows that
\begin{eqnarray}
w(z=\eta)=\eta_t+(u+U_0+\Omega \eta)_{z=\eta}\eta_x \\
w(z=\eta)=\eta_t+[u(z=\eta)+U_0+\Omega \eta]\eta_x 
\end{eqnarray}
Equation (\ref{interal-continuite}) becomes
\begin{equation}
-\int_{-h}^{\eta(x,t)} u_x dz = \eta_t+[u(z=\eta)+U_0+\Omega \eta]\eta_x 
\end{equation}
Using equation (\ref{interal-continuite-2}) with $h_x=0$ we obtain
\begin{equation}
\frac{\partial}{\partial x}\int_{-h}^{\eta(x,t)}udz + \eta_t + (U_0+\Omega \eta) \eta_x=0
\end{equation}
We assume $u$ does not depend on $z$, then
\begin{equation}
\eta_t + \frac{\partial}{\partial x}[u(\eta+h)+\frac{\Omega}{2}\eta^2 + U_0 \eta]=0
\label{mass_conservation}
\end{equation}
Equation (\ref{mass_conservation}) corresponds to mass conservation in shallow water in the presence of constant vorticity.
\vspace{0.2cm}
\newline
Under the assumption of hydrostatic pressure, the Euler equation in $x$-direction is
\begin{equation}
u_t + (u+U_0+\Omega z)u_x + \Omega w +g \eta_x=0
\end{equation}
where $g$ is the gravity.
\vspace{0.2cm}
\newline
Using the continuity equation and boundary conditions that $w$ satisfies on the bottom and at the free surface, we obtain
\begin{equation}
w=-(z+h)u_x
\end{equation}
It follows that the Euler equation becomes
\begin{equation}
u_t + (u+U_0-\Omega h)u_x + g \eta_x=0
\label{x_Euler}
\end{equation}
The dynamics of non dispersive shallow water waves on a vertically sheared current of constant vorticity is governed by equations (\ref{mass_conservation}) and (\ref{x_Euler}).
\vspace{0.2cm}
\newline
The pair of equations (\ref{mass_conservation}) and (\ref{x_Euler}) admits the following Riemann invariants 
\begin{equation}
u+\frac{\Omega H}{2} \pm \bigg\lbrace\sqrt{gH+\Omega^2H^2/4}+\frac{g}{\Omega}\ln\left[1+\frac{\Omega}{2g}(\Omega H +2 \sqrt{gH+\Omega^2H^2/4})\right]\bigg\rbrace=\rm{constant} \nonumber
\end{equation}
on characteristic lines
\begin{equation}
\frac{dx}{dt}=u+U_0+\frac{1}{2}\Omega (\eta-h)\pm \sqrt{gH+\frac{\Omega^2 H^2}{4}} 
\label{characteristic}
\end{equation}
where $H=\eta + h$.
\vspace{0.2cm}
\newline
The constant is determined for $u=0$ and $\eta = 0$ or $H=h$.
\vspace{0.2cm}
\newline
Finally 
\begin{eqnarray}
u+\frac{\Omega \eta}{2} &+& \sqrt{gH+\Omega^2H^2/4}-\sqrt{gh+\Omega^2h^2/4} \nonumber \\ &+&  \frac{g}{\Omega}
\ln\left[\frac{1+\frac{\Omega}{2g}(\Omega H
+2 \sqrt{gH+\Omega^2H^2/4})}{1+\frac{\Omega}{2g}(\Omega h
+2 \sqrt{gh+\Omega^2h^2/4})}\right]=0 
\end{eqnarray}
\begin{eqnarray}
u+\frac{\Omega \eta}{2} &-& \sqrt{gH+\Omega^2H^2/4}+\sqrt{gh+\Omega^2h^2/4} \nonumber \\ &-& \frac{g}{\Omega}
\ln\left[\frac{1+\frac{\Omega}{2g}(\Omega H
+2 \sqrt{gH+\Omega^2H^2/4})}{1+\frac{\Omega}{2g}(\Omega h
+2 \sqrt{gh+\Omega^2h^2/4})}\right]=0
\end{eqnarray}
Let us consider a wave moving rightwards
\begin{eqnarray}
u= -\frac{\Omega \eta}{2} &+& \sqrt{gH+\Omega^2H^2/4}-\sqrt{gh+\Omega^2h^2/4} \nonumber \\ &+&   \frac{g}{\Omega}
\ln\left[\frac{1+\frac{\Omega}{2g}(\Omega H
+2 \sqrt{gH+\Omega^2H^2/4})}{1+\frac{\Omega}{2g}(\Omega h
+2 \sqrt{gh+\Omega^2h^2/4})}\right]
\end{eqnarray}
Substituting this expression into equation (\ref{mass_conservation}) gives
\begin{equation}
\eta_t + \bigg\lbrace U_0-\frac{\Omega h}{2} + 2\sqrt{g(\eta+h)+\Omega^2(\eta+h)^2/4}-\sqrt{gh+\Omega^2h^2/4} 
\nonumber
\end{equation}
\begin{equation}
 + \frac{g}{\Omega} \ln\left[1+ \frac{\Omega}{2g}\frac{\Omega \eta +2(\sqrt{g(\eta+h)+\Omega^2(\eta+h)^2/4}-\sqrt{gh+\Omega^2h^2/4})}{1+\frac{\Omega}{g}(\frac{\Omega h}{2}+\sqrt{gh+\Omega^2h^2/4})} \right] \bigg\rbrace\eta_x =0 
\label{nonlin-nondispersive}
\end{equation}
Equation (\ref{nonlin-nondispersive}) is fully nonlinear and describes the spatio-temporal evolution of hyperbolic water waves in shallow water in the presence of constant vorticity. This equation is equivalent to the system of equations (\ref{mass_conservation}) and (\ref{x_Euler}) for waves moving rightwards.
\vspace{0.3cm}
\newline
Following \citet{Whitham}, full linear dispersion is introduced heuristically
\begin{equation}
\eta_t + \bigg\lbrace U_0-\frac{\Omega h}{2} + 2\sqrt{g(\eta+h)+\Omega^2(\eta+h)^2/4}-\sqrt{gh+\Omega^2h^2/4}
\nonumber 
\end{equation}
\begin{equation}
+ \frac{g}{\Omega} \ln\left[1+ \frac{\Omega}{2g}\frac{\Omega \eta +2(\sqrt{g(\eta+h)+\Omega^2(\eta+h)^2/4}-\sqrt{gh+\Omega^2h^2/4})}{1+\frac{\Omega}{g}(\frac{\Omega h}{2}+\sqrt{gh+\Omega^2h^2/4})} \right] \bigg\rbrace\eta_x + K*\eta_x=0 
\label{nonlin-dispersive}
\end{equation}
where $K*\eta_x$ is a convolution product. The kernel $K$ is given as the inverse Fourier transform of the fully linear dispersion relation of gravity waves in finite depth in the presence of constant vorticity $\Omega$: $K=F^{-1}(c)$ with 
\[c=U_0-\frac{\Omega \tanh (kh)}{2k} + \sqrt{\frac{g \tanh(kh)}{k} + \frac{\Omega^2 \tanh^2(kh)}{4 k^2}}\]
Equation (\ref{nonlin-dispersive}) governs the propagation of nonlinear long gravity waves in a fully linear dispersive medium. For $\Omega=0$ and $U_0=0$ (\ref{nonlin-nondispersive}) reduces to equation (13.97) of \citet{Whitham}. 
\vspace{0.2cm}
\newline
For weakly nonlinear waves ($\eta/h\ll1$) equation (\ref{nonlin-dispersive}) becomes the Whitham equation with constant vorticity given by 
\begin{equation}
\eta_t + \frac{3gh+h^2 \Omega^2}{h\sqrt{gh(4gh+h^2 \Omega^2)}} \eta \eta_x + K*\eta_x=0 
\label{Whitham}
\end{equation}
Note that in the Whitham equation the exact linear dispersion is considered unlike the KdV equation. In the absence of vorticity, \citet{Ehrnstrom} proved rigorously that equation (\ref{Whitham}) admits small-amplitude periodic travelling wave solutions and computed numerically approximations of small- and finite-amplitude waves as well as those close to the highest. The existence of solitary waves has been proven by \citet{Ehrnstrom1} without considering vorticity effect.
\newline
The numerical method that allows the computation of periodic travelling wave solutions of equation (\ref{Whitham}) is included in Appendix A. In figure \ref{fig:bifurcation_vor} is plotted the maximum elevation of the solutions of equation (\ref{Whitham}) for branches corresponding to several values of the vorticity. For a constant maximum elevation the phase velocity decreases as $\Omega$ increases. In figure \ref{fig:profiles_Whitham_vor} are shown the profiles of the elevation $\eta$ for several values of the vorticity for a constant phase velocity and a constant maximum elevation, respectively. The profiles become larger as $\Omega$ is decreasing (or as the vorticity is increasing).
\begin{figure}
\center
\includegraphics[width=0.6\linewidth]{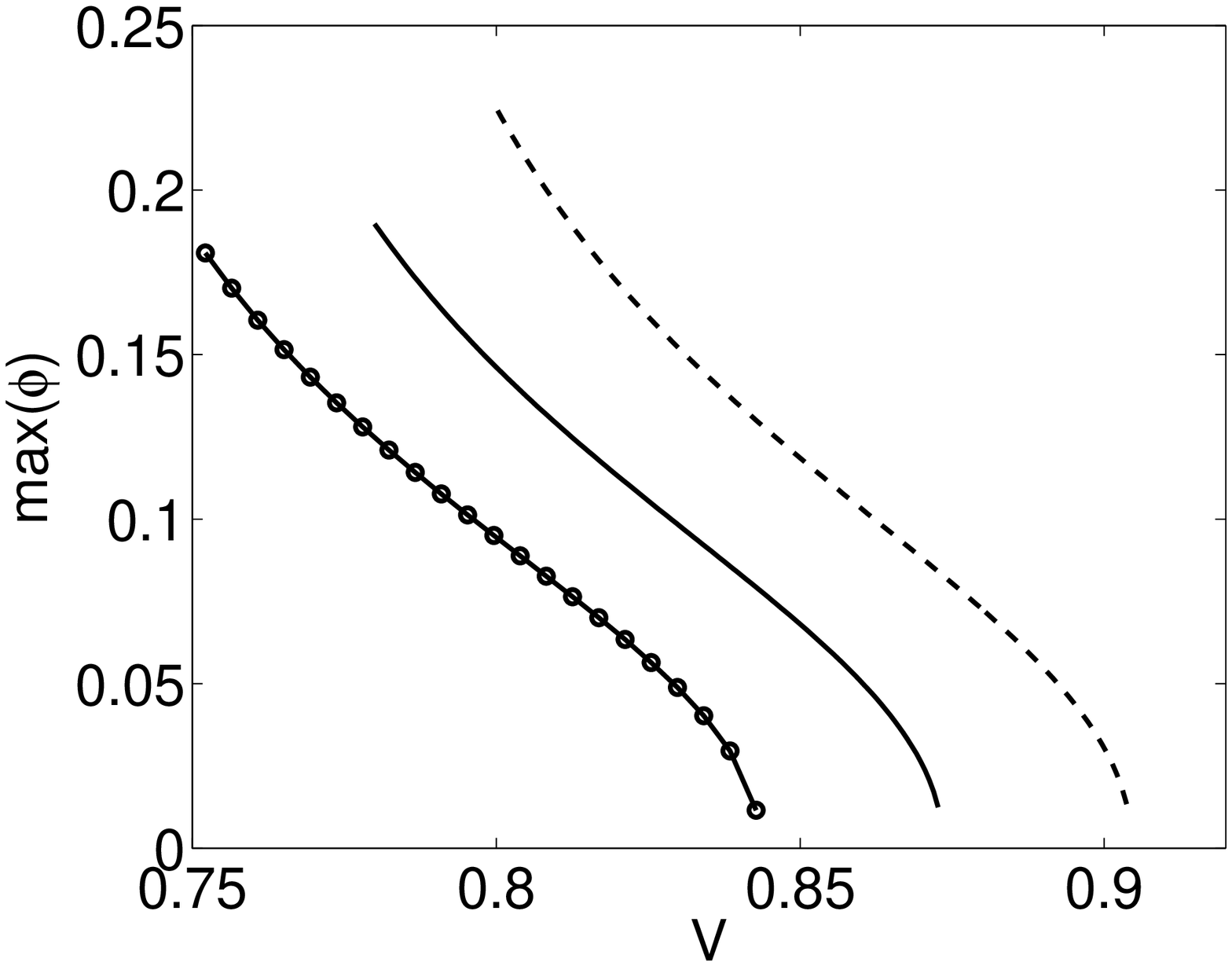}
\includegraphics[width=0.6\linewidth]{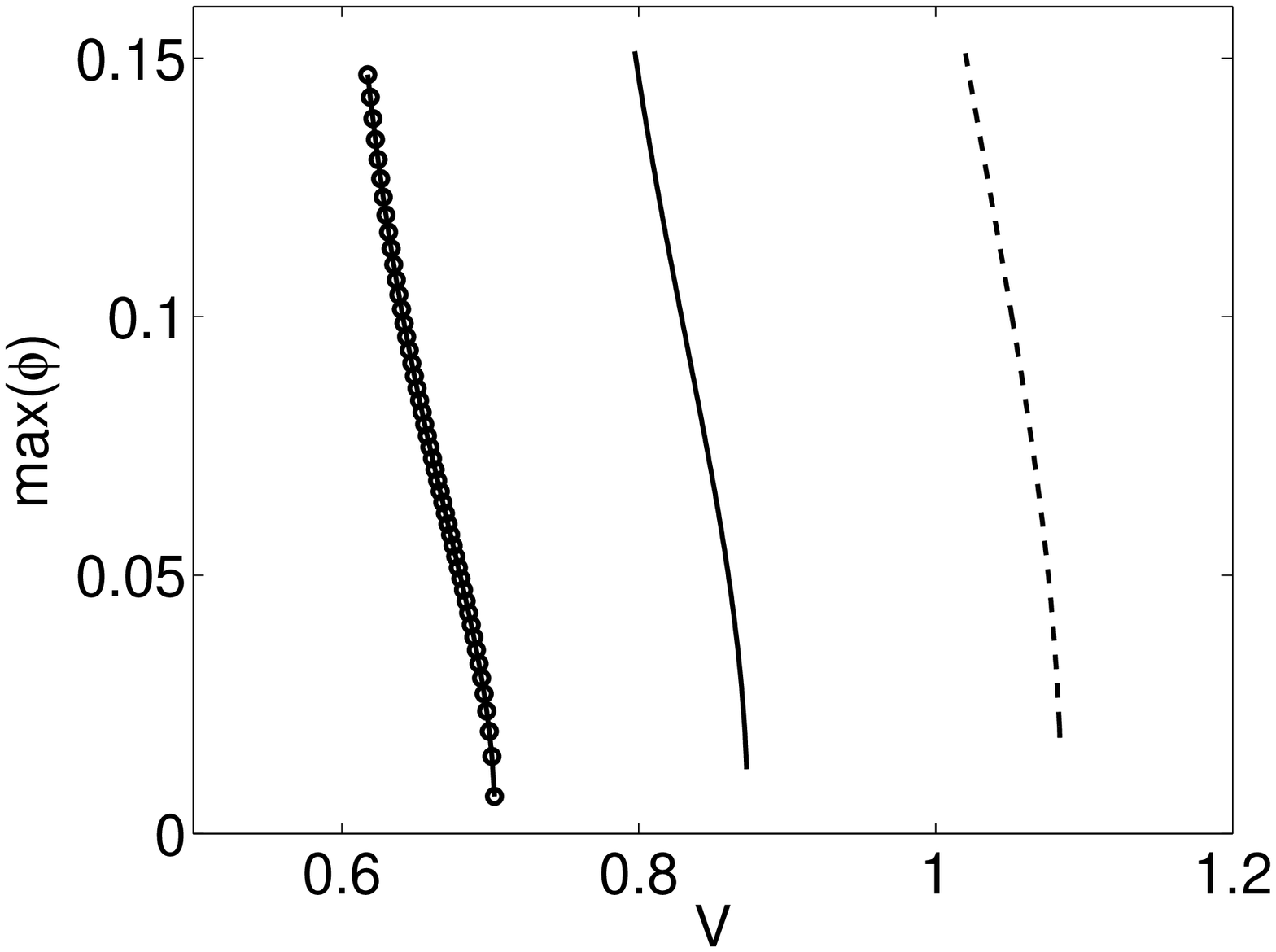}
\caption{Bifurcation diagrams for several values of the vorticity in the (maximum elevation, phase velocity) plane. Top: solid line ($\Omega=0$), dashed line ($\Omega=-0.08$) and -$\circ$- ($\Omega=0.08$). Bottom: solid line ($\Omega=0$), dashed line ($\Omega=-0.5$) and -$\circ$- ($\Omega=0.5$)}
\label{fig:bifurcation_vor}
\end{figure}
\begin{figure}
\center
\includegraphics[width=0.6\linewidth]{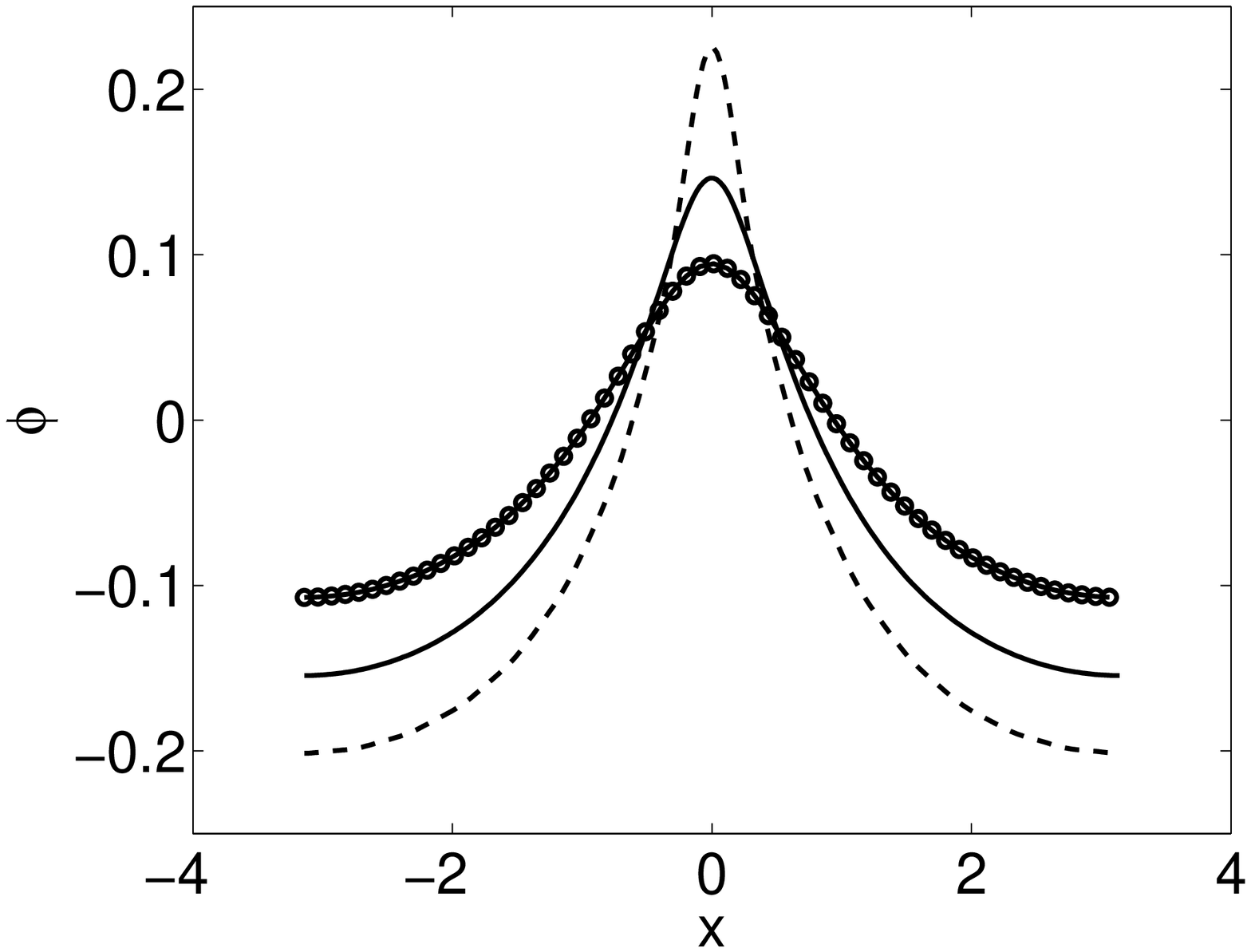}
\includegraphics[width=0.6\linewidth]{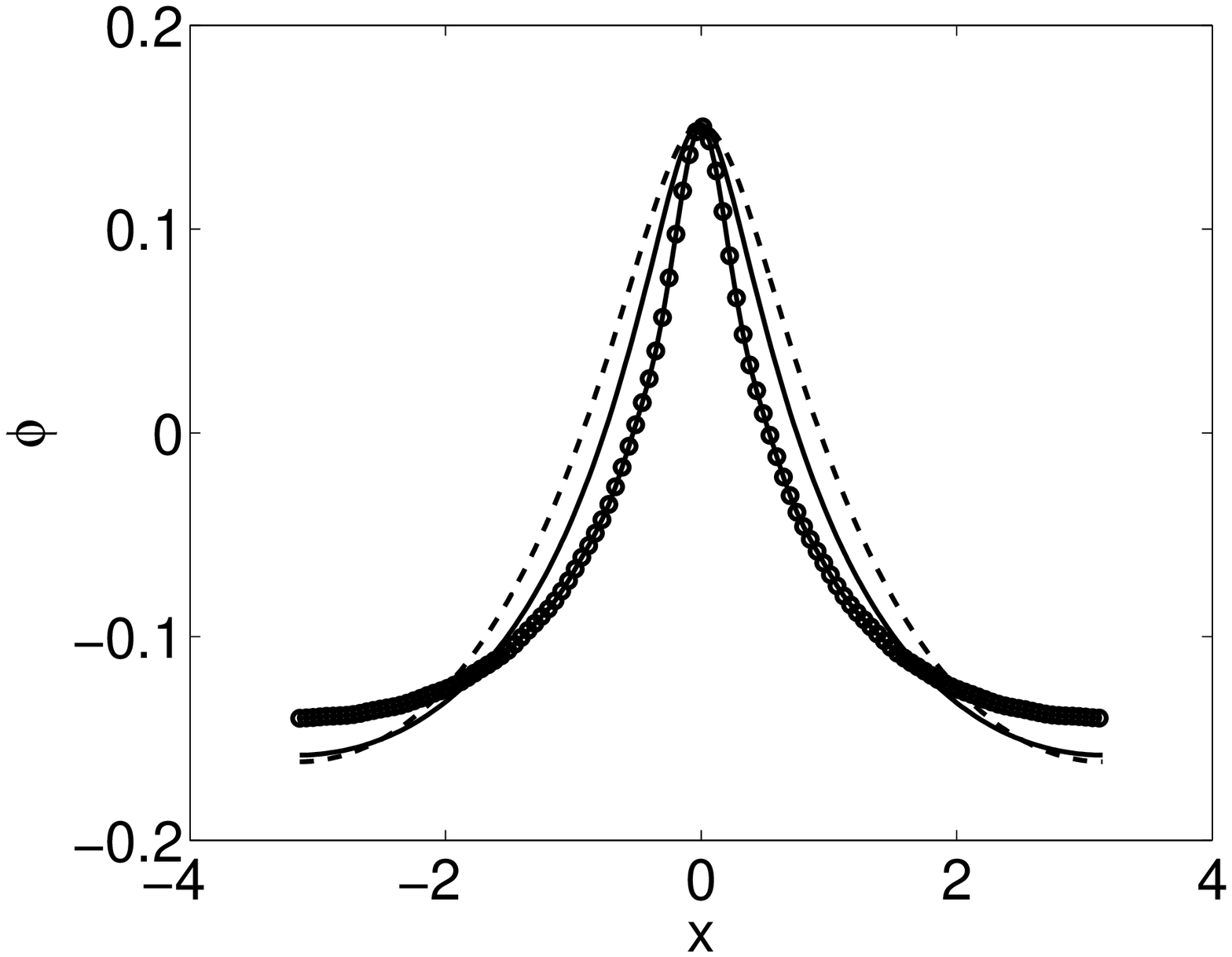}
\caption{Elevation profiles. Top: solid line ($\Omega=0$), dashed line ($\Omega=-0.08$) and -$\circ$- ($\Omega=0.08$) for $V=0.8$. Bottom: solid line ($\Omega=0$), dashed line ($\Omega=-0.5$) and -$\circ$- ($\Omega=0.5$) for the maximum of elevation equal to $0.15$}.
\label{fig:profiles_Whitham_vor}
\end{figure}
\vspace{0.2cm}
\newline
We call equation (\ref{nonlin-dispersive}) which generalises the Whitham equation (\ref{Whitham}) to fully nonlinear waves the generalised Whitham equation in the presence of vorticity.
\vspace{0.1cm}
\newline
For weakly nonlinear ($\eta/h\ll1$) and weakly dispersive ($kh\ll1$) water waves, equation (\ref{nonlin-dispersive}) reduces to the KdV equation with vorticity derived by \citet{Freeman}  and \citet{Choi} who used multiple scale methods, different to the approach used herein. To set the KdV equation in dimensionless form, $h$ and $\sqrt{h/g}$ are chosen as reference length and reference time which corresponds to $h=1$ and $g=1$. The equation reads 
\begin{equation}
\eta_t+c_0(\Omega) \eta_x+c_1(\Omega)\eta \eta_x+c_2(\Omega)\eta_{xxx}=0
\label{vor-KdV}
\end{equation}
with
\[c_0=U_0-\frac{\Omega}{2}+\sqrt{1+\Omega^2/4}, \quad\, \quad c_1=\frac{3+\Omega^2}{\sqrt{4+\Omega^2}}, \quad   \quad c_2=\frac{2+\Omega^2-\Omega\sqrt{4+\Omega^2}}{6\sqrt{4+\Omega^2}}\]
Equation (\ref{vor-KdV}) is known to admit as solutions the solitary wave and cnoidal wave whose expressions are
\[\eta=a\, \mathrm{sech}^2 (\frac{x-ct}{\Delta_s})\]
with $c=c_0 + \frac{c_1}{3}a$ and $\Delta_s=\sqrt{\frac{12 c_2}{c_1 a}}$
\newline
and
\[\eta=\frac{a}{m}(1-m-\frac{E(m)}{K(m)}) + a\, \mathrm{cn}^2(\frac{(x-ct)}{\Delta_{cn}}|m) \]
with $\Delta_{cn} = \sqrt{12 m c_2/(c_1 a)}$ and $c=c_0 + c_1(2-m-E(m)/K(m))a/6$
\newline
Note that $\Delta_s=\Delta_{cn}(m=1)$. An increase of $\Delta_s$ and $\Delta_{cn}$ gives a wider wave profile. The profiles of the solitary wave and cnoidal wave in the presence of vorticity ($-\Omega$) for various values of $\Omega$ are plotted in figure \ref{fig:profiles:waves}. In both cases, for fixed height $a$ the width of the profiles is increasing with the vorticity. In figure \ref{fig:width_vor} is shown $\sqrt{a/12}\Delta_{cn}$ as a function of $\Omega$ for two values of the elliptic parameter $m=1$ and $m=0.5$. Solitary waves and cnoidal waves of height $a$ propagating on currents of positive vorticity ($\Omega<0$) are wider than solitary waves and cnoidal waves of height $a$ propagating on currents of negative vorticity ($\Omega>0$).  
\begin{figure}
\includegraphics[width=0.5\linewidth]{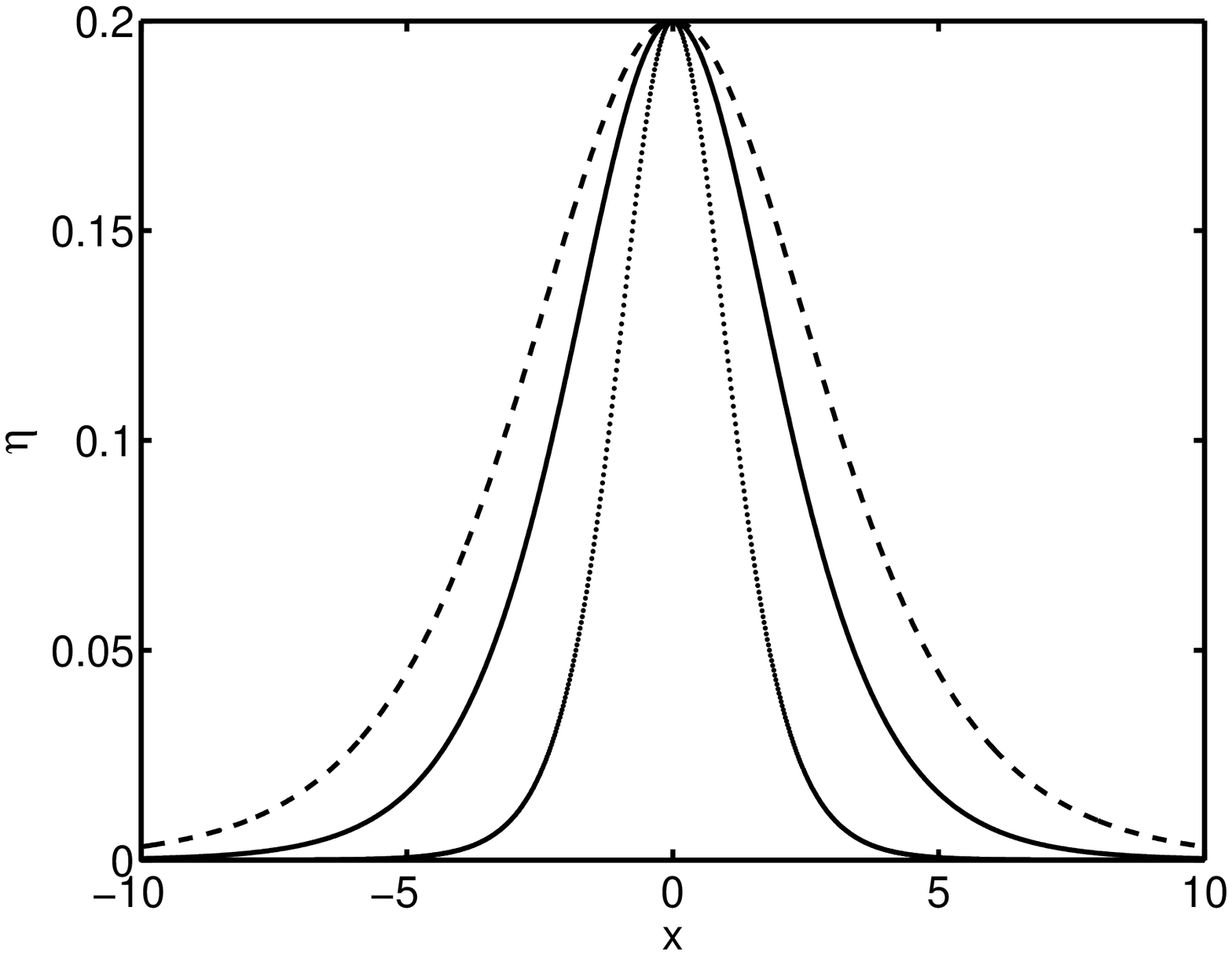}
\includegraphics[width=0.5\linewidth]{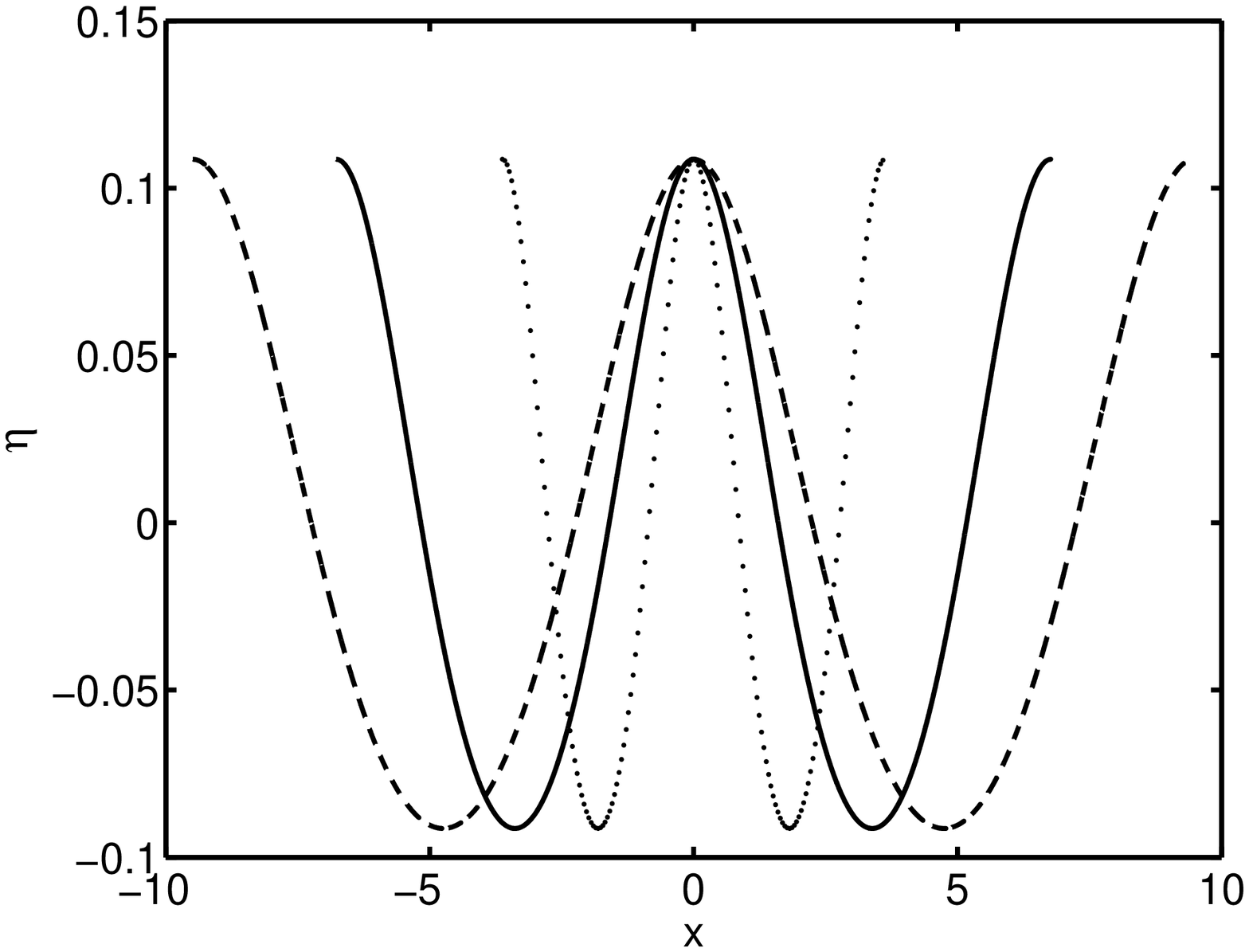}
\caption{Profiles of solitary waves and cnoidal waves for various values of the vorticity. Solid line ($\Omega=0$), dashed line ($\Omega=-1$) and dotted line ($\Omega=1$)}
\label{fig:profiles:waves}
\end{figure}
\begin{figure}
\center
\includegraphics[width=0.7\linewidth]{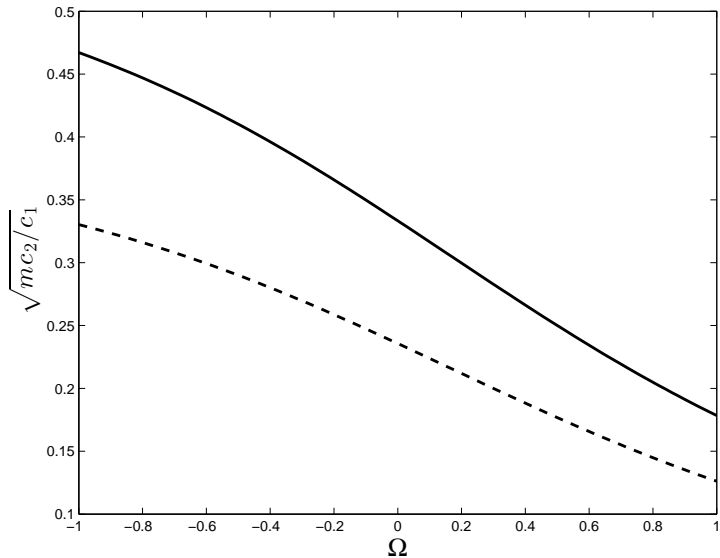}
\caption{Width of the profiles of solitary waves (solid line for $m=1$) and cnoidal waves (dashed line for $m=0.5$) as a function of the vorticity}
\label{fig:width_vor}
\end{figure}
\vspace{0.2cm} 
\newline
The dimensionless form of the Whitham equation (\ref{Whitham}) is
\begin{equation}
\eta_t+c_1(\Omega)\eta \eta_x+K*\eta_x=0
\end{equation}
with
\[c^*=U_0-\frac{\Omega \tanh k}{2k} + \sqrt{\tanh k(1+\Omega^2 \tanh k/(4k))/k}, \quad c_1=\frac{3+\Omega^2}{\sqrt{4+\Omega^2}}\]
Herein $K=F^{-1}(c^*)$

\section{Numerical integration}
The equations (\ref{nonlin-nondispersive}), (\ref{nonlin-dispersive}) and (\ref{vor-KdV}) are solved numerically in a periodic domain of length $2L$. The length $L$ is chosen $O(400\delta)$ where $\delta$ is a characteristic length scale of the initial condition. The number of grid points is $N_x=2^{14}$. Spatial derivatives are computed in the Fourier space and nonlinear terms in the physical space. The link between the two spaces is made by the Fast Fourier Transform.

\begin{figure}
\includegraphics[width=0.3\linewidth]{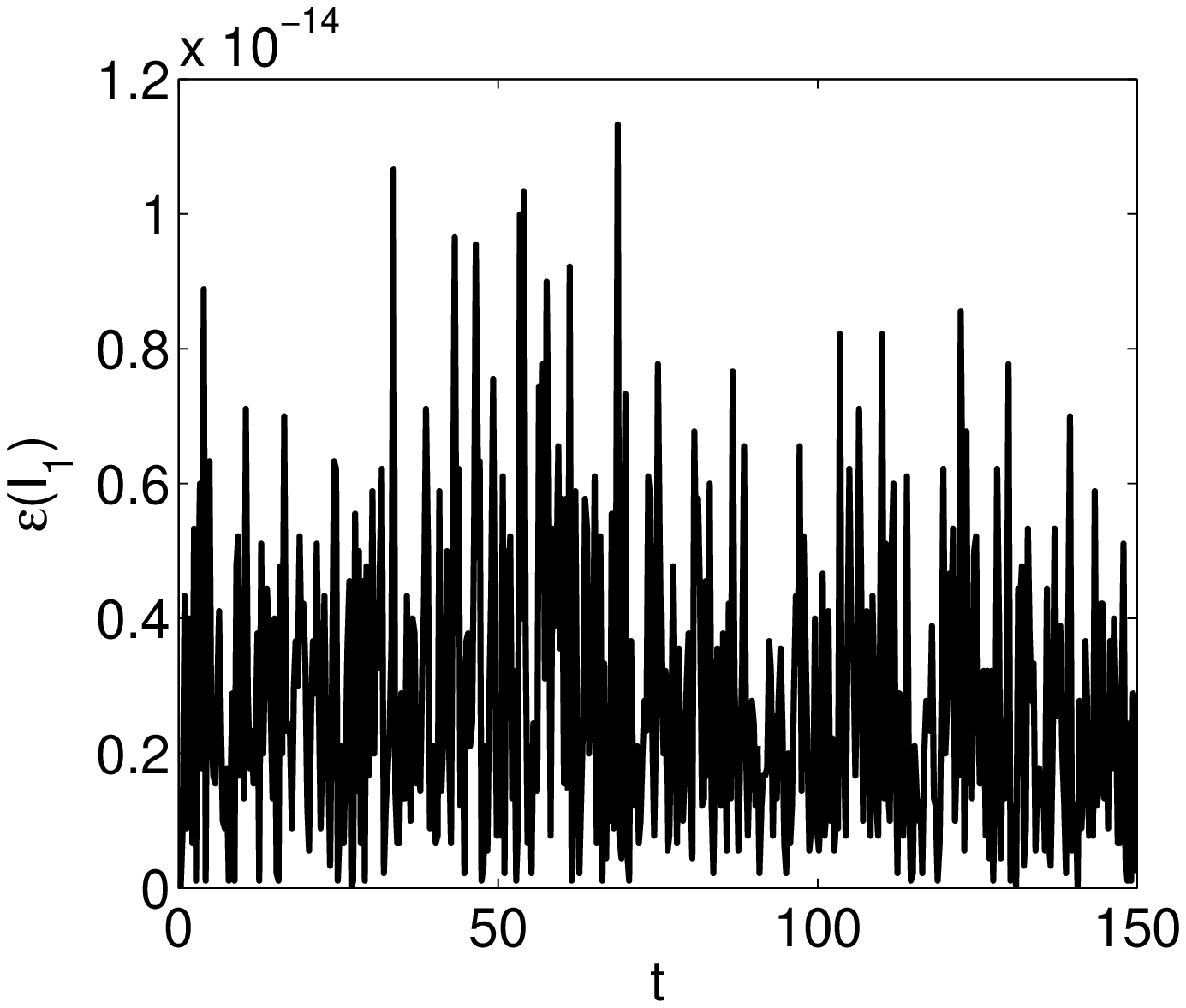}
\includegraphics[width=0.3\linewidth]{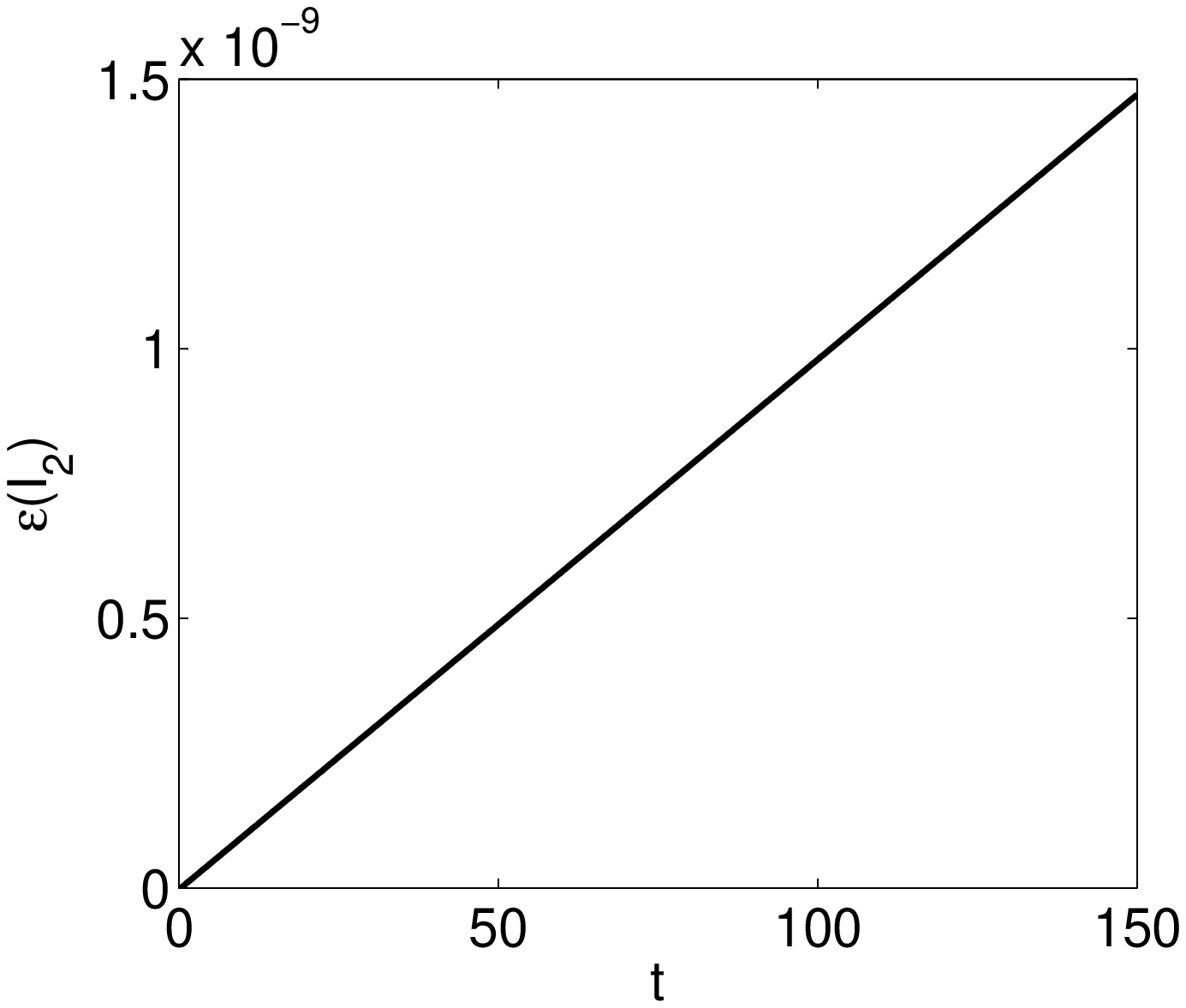}
\includegraphics[width=0.3\linewidth]{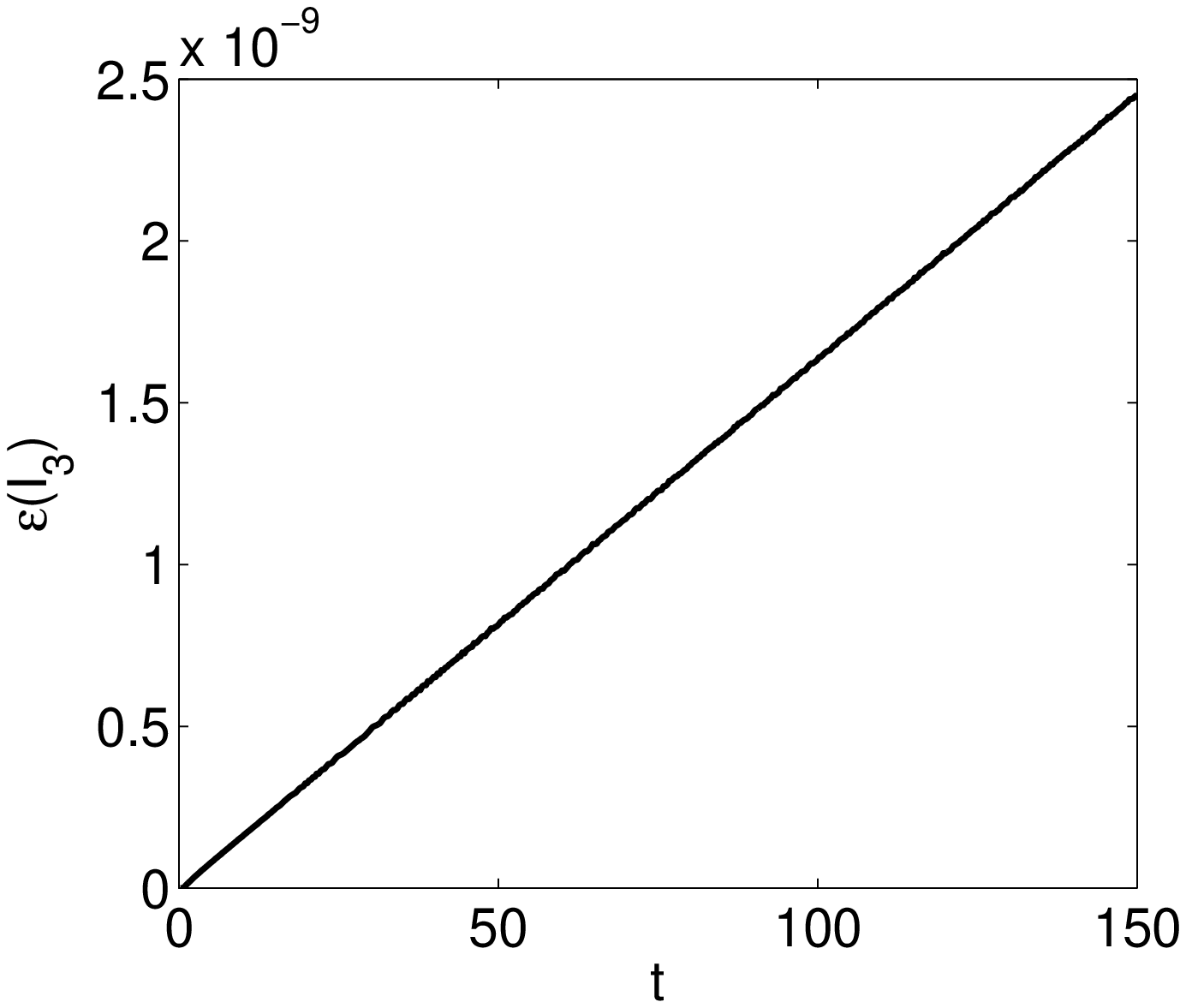}
\caption{Relative errors of the computed invariants}
\label{fig:rel:error}
\end{figure}

For the time integration, a splitting technique is used. The equations (\ref{nonlin-nondispersive}), (\ref{nonlin-dispersive}) and (\ref{vor-KdV}) could be written as
\begin{equation}
\eta_t+L+N=0,
\end{equation}
where $L$ and $N$ are linear and nonlinear differential operators in $\eta$, respectively. Note that in general the operators $L$ and $N$ do not commute. If the initial condition is $\eta_0$, the exact solution of the previous equation is 
\begin{equation}
\eta(t)=e^{-(L+N)t}\eta_0.
\end{equation}
This equation is discretized as follows. Let $t_n=n\Delta t$. We have
\begin{equation}
\eta(t_n)=e^{-(L+N)n\Delta t}\eta_0=(e^{-L\Delta t/2}e^{-N\Delta t}e^{-L\Delta t/2})^n\eta_0+O(\Delta t^2),
\end{equation}
and the scheme is globally second order in time. The operator $e^{-L\Delta t/2}$ is computed exactly in the Fourier space. However, the operator $e^{-N\Delta t}$ is approximated using a Runge-Kutta scheme of order 4. The time step is chosen as $\Delta t=0.005$. With this time step, and when the KdV equation is integrated, the mass $I_1$, momentum $I_2$ and energy $I_3$ invariants are conserved to the machine precision for $I_1$ and with a relative error $\epsilon(I_i)=(\tilde I_i(t)-I_i)/I_i=O(10^{-9})$ for $I_2$ and $I_3$ ($\tilde I_i$ being the computed value of $I_i$) as shown in figure \ref{fig:rel:error}.

\section{Validation of the numerical method}

\begin{figure}
\center
\includegraphics[width=0.7\linewidth]{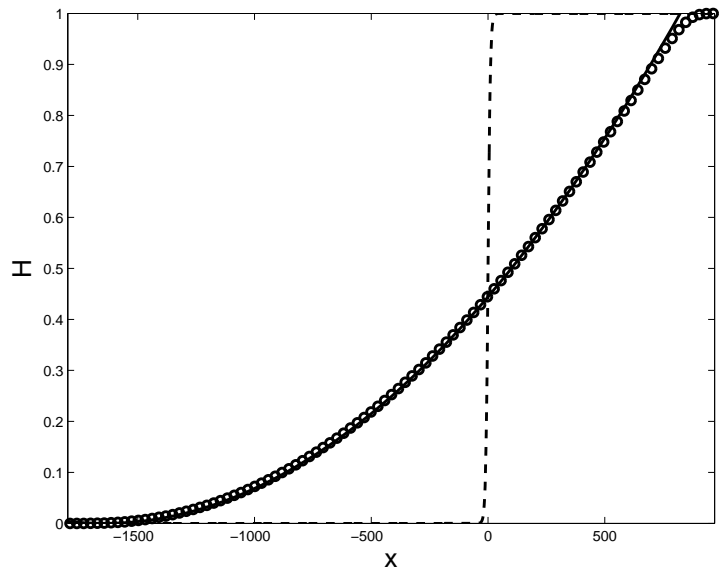}
\caption{Dam-break: comparison between analytical (solid line) and numerical solutions ($\circ$) after the dam has broken. The dashed line represents the initial condition at $t=0$ }
\label{fig:dam_break}
\end{figure}

\begin{figure}
\vspace{0.2cm}
\center
\includegraphics[width=0.6\linewidth]{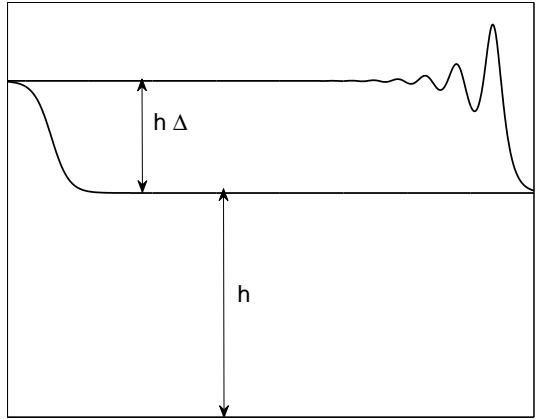}
\caption{Sketch of the evolution of an undular bore from its initial position}
\label{fig:undular_sketch}
\end{figure}

\begin{figure}
\vspace{0.2cm}
\center
\includegraphics[width=0.4\linewidth]{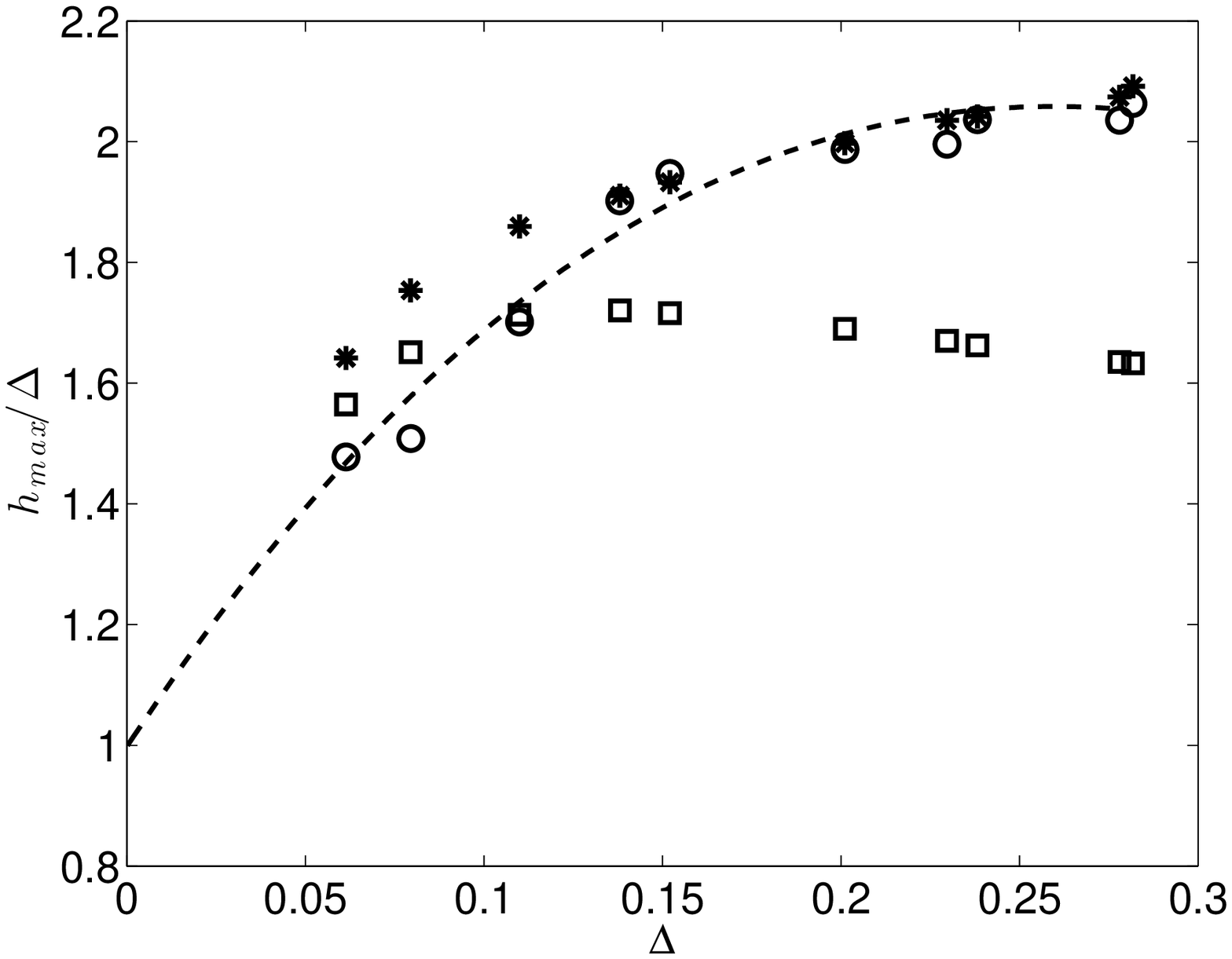}
\includegraphics[width=0.4\linewidth]{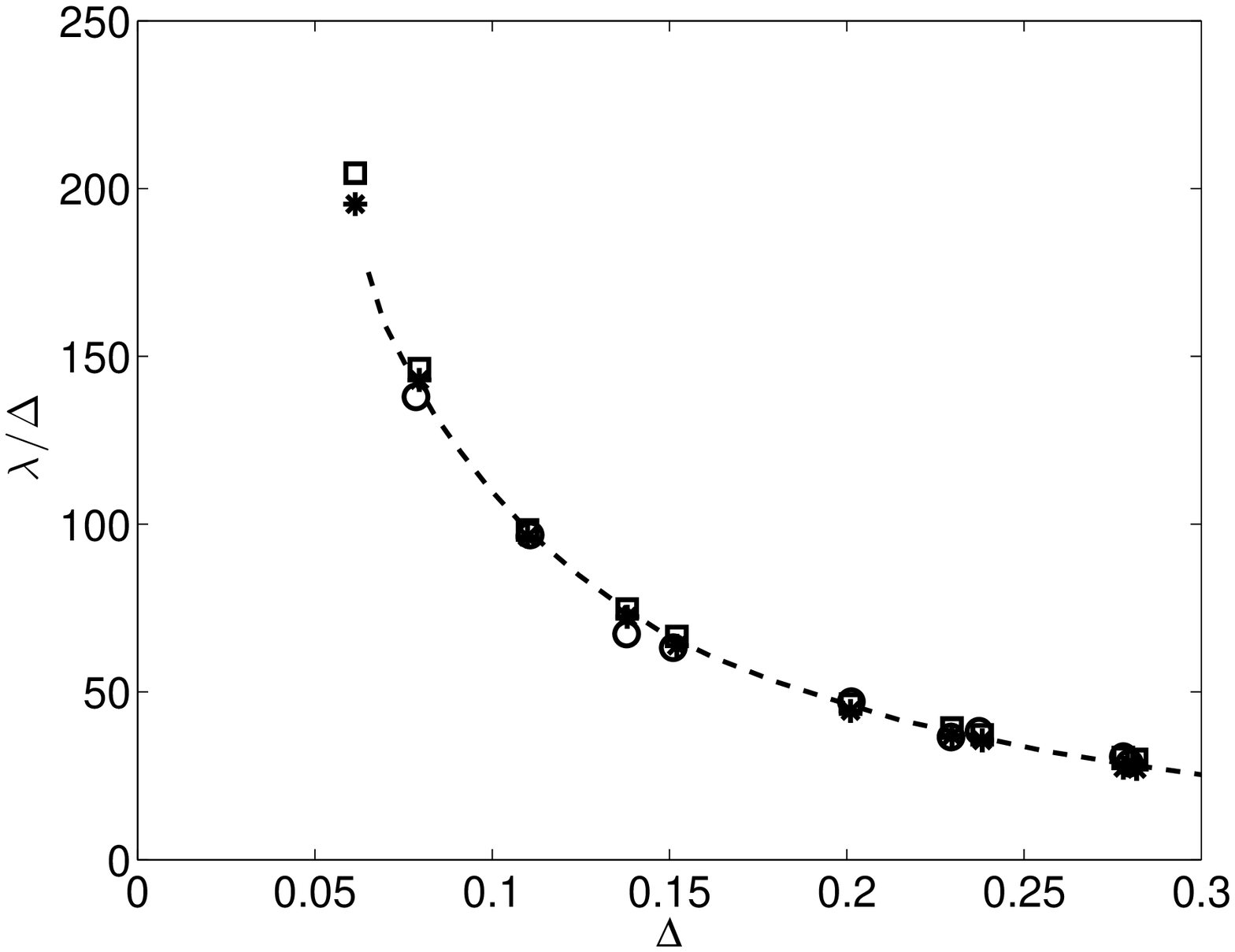}
\caption{Dimensionless height of the leading wave (left) and dimensionless wavelength of the trailing waves (right) as a function of the initial relative difference in water level. Favre's experiments ($\circ$), equation (\ref{nonlin-dispersive}) with damping and $(U_0,\Omega)=(0,0)$  ($\ast$), KdV equation with damping and $(U_0,\Omega)=(0,0)$ ($\square$) }
\label{fig:undular_comparison_Favre}
\end{figure}

The efficiency and accuracy of the numerical method has been checked against the nonlinear analytical solution of the St-Venant equations for the dam-break problem and the experiments of \citet{Favre}, in the absence of current and vorticity ($\Omega=0$ and $U_0=0$). 
\vspace{0.2cm}
\newline
For $U_0=0$ and $\Omega=0$ equation (\ref{nonlin-nondispersive}) reduces to
\begin{equation}
H_t +(3\sqrt{gH}-2\sqrt{gh})H_x=0, \quad \mathrm{with} \quad H=\eta+h. 
\label{dam-break}
\end{equation}
For $t>0$, the nonlinear analytical solution of equation (\ref{dam-break}) is
\begin{eqnarray}
H(x,t)=h, & u(x,t)=0; & \frac{x}{t}\ge\sqrt{gh} \nonumber \\
H(x,t)=\frac{h}{9}\left(2+\frac{x}{\sqrt{gh}\ t}\right)^2, & u(x,t)=-\frac{2}{3}\left(\sqrt{gh}-\frac{x}{t}\right); & -2\sqrt{gh}\le\frac{x}{t}\le\sqrt{gh} \nonumber \\
H(x,t)=0,& u(x,t)=0; & \frac{x}{t} \le -2\sqrt{gh}
\label{exact-solution}
\end{eqnarray}
At time $t=0$ the initial condition is $H(x,0)=h(1+\tanh(2x))/2$ and $u(x,0)=0$ everywhere. A numerical simulation of equation (\ref{dam-break}) has been carried out with $g=1$ and $h=1$. The numerical and analytical surface profiles at $t=0$ and after the dam has broken are plotted in figure \ref{fig:dam_break}. 
\newline
Within the framework of the KdV equation in the presence of vorticity, we have also checked that solitary waves are propagated with the right velocity that depends on $\Omega$. 
\vspace{0.2cm}
\newline
An undular bore is formed when a sudden discharge of water at rest of depth $h(1+\Delta)$ is initiated into a still water of depth $h$ (see figure \ref{fig:undular_sketch}). Herein, the bore is the region of transition between two uniform depths. To consider non-breaking undular bore, the initial relative difference in water level, $\Delta$, is chosen less than $0.28$. \citet{Favre} showed experimentally that beyond this value undular bores evolve to breaking. He conducted experiments on undular bores in a facility $73.58 \, m$ long, $0.42 \, m$ wide and $0.40 \, m$ high. He carried out a series of experiments on non breaking and breaking undular bores. Non breaking undular bores correspond to his experiments $N^{\circ}$ $2,\, 4, \, 6, \, 8, \, 10, \, 12$ for a water depth $h=0.205 \, m$ and $21, \, 22, \, 23$ for a water depth $h=0.1075 \, m$. The height of the leading wave, $h_{\mathrm{max}}$, which corresponds to a maximum waveheight was recorded at the end of the tank, after travelling on distances close to $300$ and $600$ depths. During these distances of propagation we can not ignore the shear stress acting at the bottom. Following \citet{Holloway} and \citet{Caputo} we introduce the following approximate dimensionless damping term based on the drag law for modelling the stress at the bottom
\begin{equation}
D(\eta)=k \eta |\eta|
\label{damping}
\end{equation}
where $k \sim 0.001 - 0.0026$ is an empirical dimensionless coefficient describing frictional effects at the bottom.
\newline
The time of diffusion, $t_d$, of negative vorticity generated at the bottom is $\mathcal{O}(h^2/\nu)$ where $\nu$ is the kinematic viscosity of water whereas the time of propagation, $t_p$, of the bore is $\mathcal{O}(L/\sqrt{gh})$ where $L$ is the length of the tank. For water, $t_p \ll t_d$ so that the vorticity due to molecular viscosity is assumed to have a negligible effect on the waves during their propagation. 
\newline
Let $\zeta(x,t)=\eta(x,t)-\eta_0(x)$ where $\eta_0(x)=\eta(x,0)$ is the initial condition. We substitute $\eta=\zeta+\eta_0$, $h=1$, $g=1$ and $U_0=0$ into equations (\ref{nonlin-dispersive}) and (\ref{vor-KdV}), so these equations are dimensionless. The initial condition is 
$\eta_0(x)=\Delta(1-\tanh(\alpha x))/2$ with $\alpha=1.25$, 
so that $\lim \zeta = 0, \, x \rightarrow \pm \infty$.
\newline
To compare our numerical results with those experimental of \citet{Favre} we introduce in equations (\ref{nonlin-dispersive}) and (\ref{vor-KdV}) in dimensionless form the damping term given by equation (\ref{damping}) and set $U_0=0$ and $\Omega=0$. We have neglected vortical effects due to an underlying current because experimental bores considered herein were generated in water at rest. Experimental and numerical results are shown in figure \ref{fig:undular_comparison_Favre}. The dashed line corresponds to the experimental fit obtained by Favre in his figure 49. For strongly nonlinear waves ($0.30 \lesssim h_{\mathrm{max}} \lesssim 0.60$) results given by equation (\ref{nonlin-dispersive}) are close to those of experiments whereas KdV's results are in quite good agreement with experiments for weakly nonlinear waves ($h_{\mathrm{max}} \lesssim 0.20$). Beyond $h_{\mathrm{max}}=0.20$, the Boussinesq regime is no longer valid because the Ursell number becomes large and equation (\ref{nonlin-dispersive}) with its full nonlinearity is more appropriate. Within the framework of the KdV equation, we found that the profiles of the leading wave at $300$ depths are very close to those of solitary waves.

\section{Vorticity effect on undular bore properties}
 
\begin{figure}
\vspace{0.2cm}
\center
\includegraphics[width=0.7\linewidth]{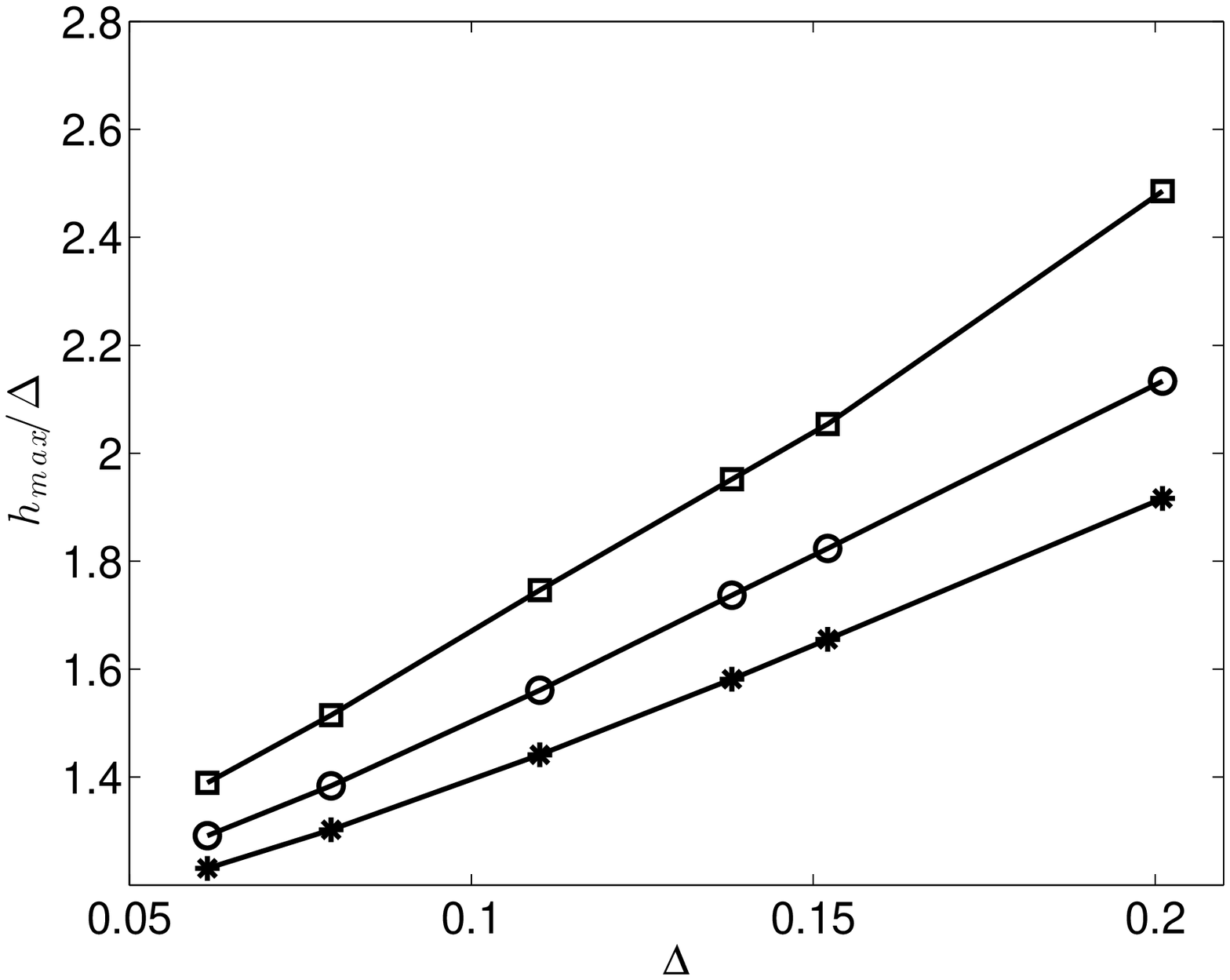}
\caption{Dimensionless height of the leading wave as a function of the initial relative difference in water level for several values of the vorticity. $\Omega=0.20$ ($\square$), $\Omega=0$ ($\circ$) and $\Omega=-0.20$ ($\ast$).}
\label{fig:undular_vor_Riemann_hmax}
\end{figure}

\begin{figure}
\vspace{0.2cm}
\center
\includegraphics[width=0.7\linewidth]{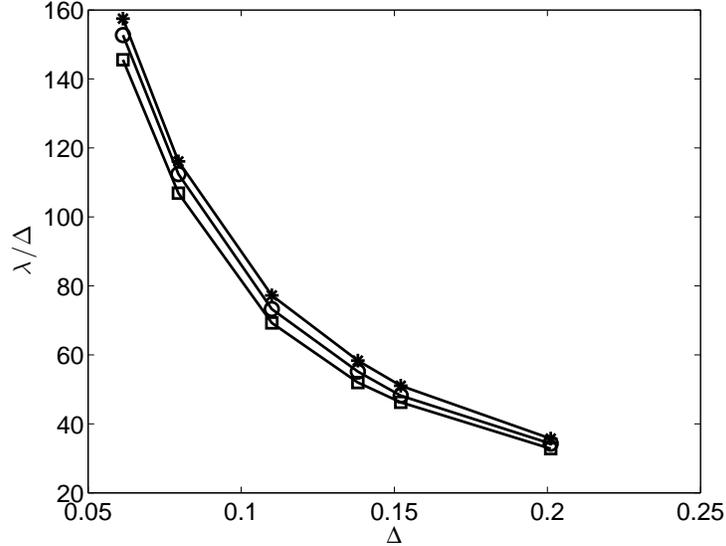}
\caption{Dimensionless wavelength of the following waves as a function of the initial relative difference in water level for several values of the vorticity. $\Omega=0.20$ ($\square$), $\Omega=0$ ($\circ$) and $\Omega=-0.20$ ($\ast$).}
\label{fig:undular_vor_Riemann_lambda}
\end{figure}

We have ignored vorticity effects due to an underlying current in our numerical simulations of Favre's experiments. Generally, in natural conditions waves travel in the presence of currents and we cannot ignore the influence of vorticity. As emphasized by \citet{Teles}  undular bores which travel upstream estuaries feel positive vorticity due to the boundary layer of the downstream current. To determine the kinematical and dynamical effects of the vorticity on the maximal height of the head wave and wavelength of the following wave train of the undular bore, numerical simulations haved been run for several values of $\Omega$. Waveheights are recorded at $100$ depths. In figure \ref{fig:undular_vor_Riemann_hmax} is shown the dimensionless height of the leading wave, $h_{\mathrm{max}}/\Delta$, as a function of the initial relative difference in water level, $\Delta$, for several values of the vorticity. In comparison to the case without vorticity and for a fixed value of $\Delta$, negative vorticity or positive vorticity increases or decreases the height of the leading wave, respectively. In figure \ref{fig:undular_vor_Riemann_lambda} is plotted the wavelength of the following waves as a function of $\Delta$ for several values of the vorticity. For a fixed value of $\Delta$, positive vorticity shortens the wavelength whereas negative vorticity lengthens the wavelength.
\newline

\section{Breaking time for hyperbolic waves and dispersive waves in the presence of vorticity}

\subsection{Within the framework of the hyperbolic equation (\ref{nonlin-nondispersive})}

Herein, we consider equation (\ref{nonlin-nondispersive}) in its dimensionless form by setting $g=1$ and $h=1$. Let us assume that $U_0=0$. 
\newline
It is well known that nonlinear water waves in shallow water may evolve to breaking. The problem of breaking waves within the framework of nonlinear hyperbolic system has been tackled by many authors. The corresponding theory for the computation of the breaking time can be found in the book of \citet{Whitham}. For most water wave models in shallow water, wave breaking corresponds to the occurrence of an infinite slope of the wave profile. 
\newline
The dimensionless form of equation (\ref{nonlin-nondispersive}) can be rewritten as follows
\begin{equation}
\eta_t+ \mathcal{C}(\eta) \eta_x=0
\label{hyperbolic-eq}
\end{equation}
where $\eta$ is now dimensionless and 
\[\mathcal{C}(\eta)=-\frac{\Omega}{2}+ 2\sqrt{(\eta+1)+\Omega^2(\eta+1)^2/4}\]
\[ -\sqrt{1+\Omega^2/4} + \frac{1}{\Omega} \ln\left[1+ \frac{\Omega}{2}\frac{\Omega \eta +2(\sqrt{(\eta+1)+\Omega^2(\eta+1)^2/4}-\sqrt{1+\Omega^2/4})}{1+\Omega(\frac{\Omega}{2}+\sqrt{1+\Omega^2/4})} \right]\]
Equation (\ref{hyperbolic-eq}) is equivalent to the following system
\[\frac{d\eta}{dt}=0, \quad \mathrm{along \, the \, characteristic \, curve} \quad
 \frac{dx}{dt}=\mathcal{C}(\eta)\]
The characteristic curves are straight lines in the $(x,t)$-plane.
\vspace{0.2cm}
\newline
Let $\eta_0(x)=\eta(x,0)$ be the initial condition and $x_0$ the point where the characteristic curve intersect the $x$-axis ($t=0$). The equation of this characteristic curve is
\[x=x_0+\mathcal{C}(\eta_0(x_0))t \]
\[x=x_0+\mathcal{V}(x_0)t \]
One can easily demonstrate that the slope of the profile at $t$ is 
\[\frac{\partial{\eta}}{\partial x}=\frac{d\eta_0/dx_0}{1+\frac{d\mathcal{V}}{dx_0}t}\]
On any charateristic for which $\frac{d\mathcal{V}}{dx_0} < 0$ the slope of the profile becomes infinite when $t=-(d\mathcal{V}/dx_0)^{-1}$.
Consequently, breaking wave first occurs on the characteristic curve intersecting the $x$-axis at $x_0=x_{0_B}$ for which $\frac{d\mathcal{V}}{dx_0}(x_{0_B}) < 0$ with $|\frac{d\mathcal{V}}{dx_0}(x_{0_B})|$ is a maximum. The breaking time is
\begin{equation}
t_B=-(\frac{d\mathcal{V}}{dx_0}(x_{0_B}))^{-1}
\label{breaking_time}
\end{equation}
Herein, the breaking wave phenomenon can be understood as the blow-up of the slope in finite time $t_B$.
\vspace{0.2cm}
\newline
For dispersive waves in shallow water, there is no analytical expression of the breaking time similar to equation (\ref{breaking_time}). Nevertheless, the determination of the breaking time can be carried out numerically. To that purpose we have tested the efficiency of a numerical method for the detection of blow-up in finite time against the expression given by (\ref{breaking_time}). The chosen numerical method which allows the detection of possible occurrence of blow-up in finite time for a large class of nonlinear evolution equations was proposed by \citet{Sulem83}. The method is briefly presented in Appendix B. In figure \ref{fig:breaking_time_hyperbolic} is plotted the breaking time as a function of the vorticity. The initial condition is 
\begin{equation}
\eta(x,0)=a \cos(kx)+ \frac{3-\sigma^2}{4\sigma^3}a^2 k \cos(2kx+\varphi)
\label{wind_profile}
\end{equation}
where $\sigma=\tanh(kh)$, $h=1$, $k=1$, $a=0.10$ and $\varphi = 0$.
\newline
Note that for $\varphi=0$ the initial profile is symmetric whereas for $\varphi \neq 0$ it is asymmetric. 
In the latter case the initial asymmetric profile is typical of wind wave profiles with the leeward side steeper than the backward side.
\newline
The agreement between the analytical and numerical results is excellent. We can observe that the breaking time decreases as the current intensity increases in agreement with the results of Hur (2017). The presence of the shear current reduces the breaking time and speeds up the breaking phenomenon.
\begin{figure}
\vspace{0.2cm}
\center
\includegraphics[width=0.7\linewidth]{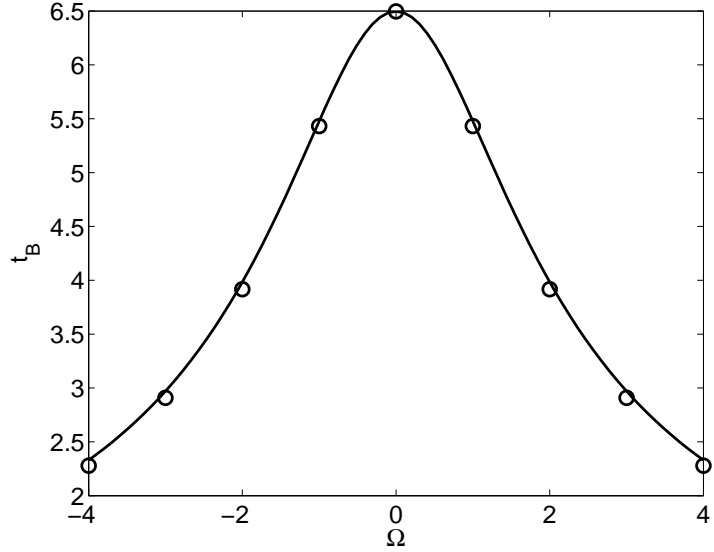}
\caption{Theoretical and numerical breaking times as a function of the vorticity within the framework of equation (\ref{nonlin-nondispersive}). The solid line corresponds to the theoritical solution whereas the circles correspond to numerical values.}
\label{fig:breaking_time_hyperbolic}
\end{figure}

\subsection{Within the framework of the Whitham equation}

\begin{figure}
\vspace{0.2cm}
\center
\includegraphics[width=0.6\linewidth]{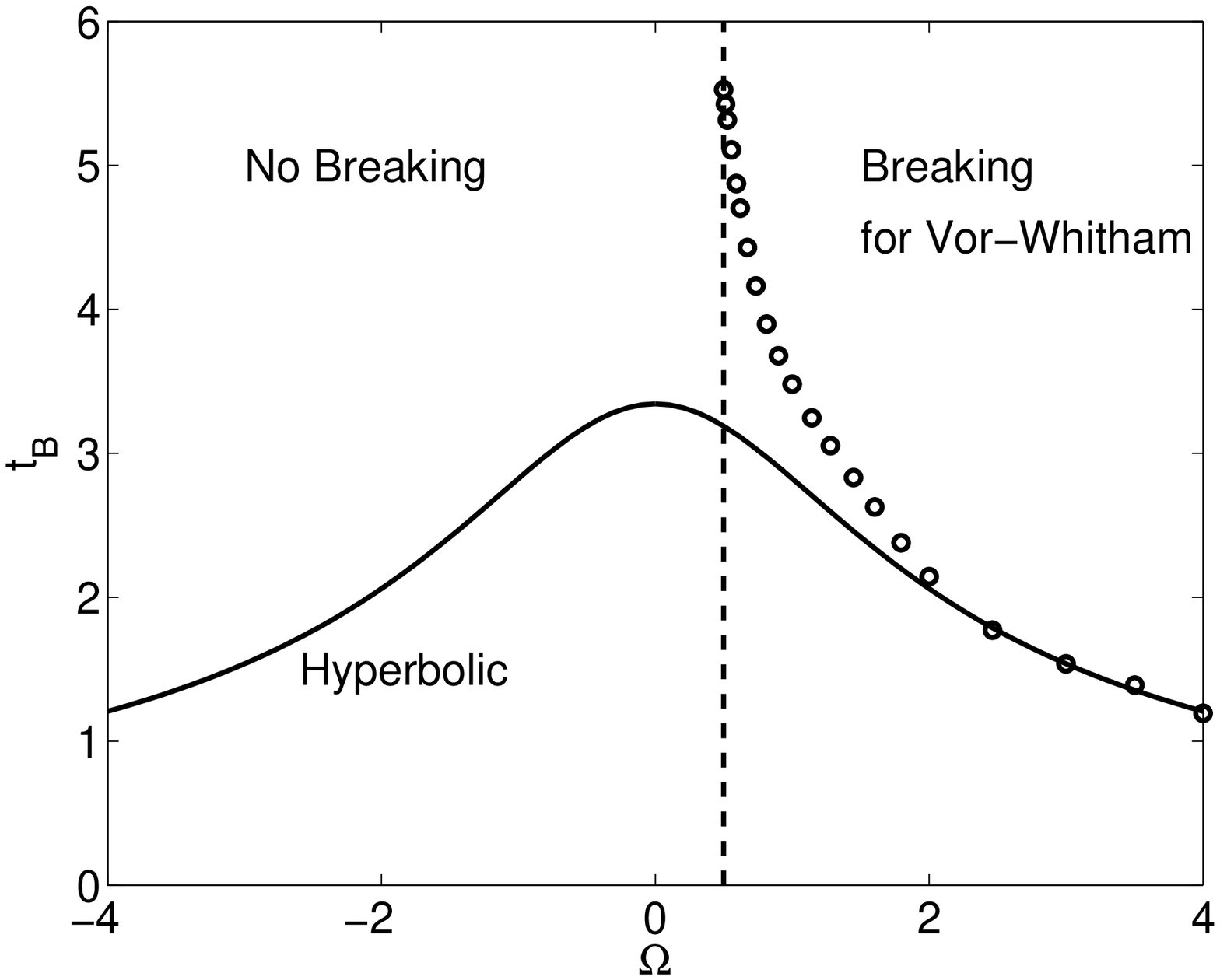}
\includegraphics[width=0.6\linewidth]{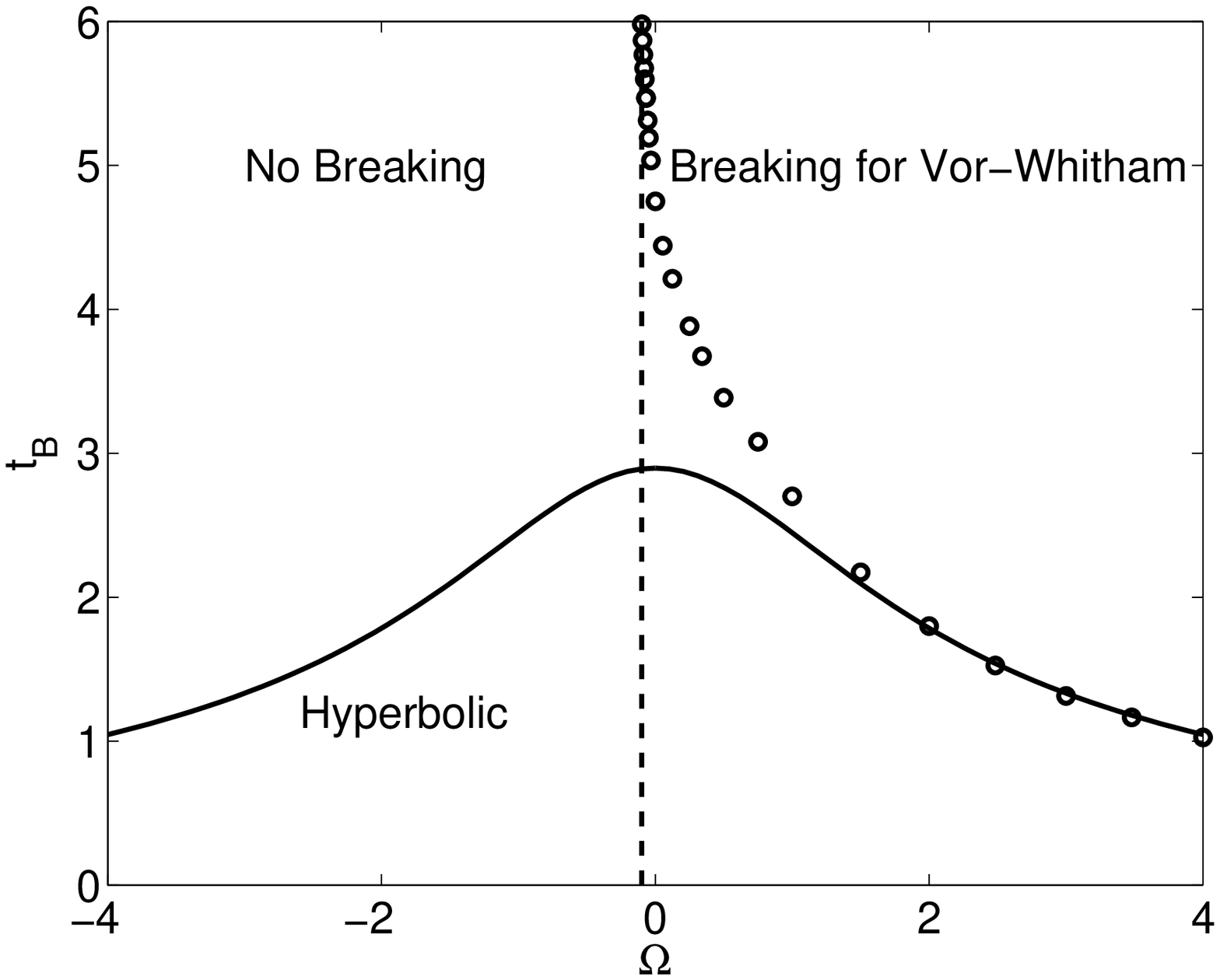}
\caption{Breaking time as a function of the vorticity for the symmetric initial condition ($a=0.16, \,\varphi=0$) (top) and for the asymmetric initial condition ($a=0.16, \,\varphi=3\pi/2$) (bottom). The solid line corresponds to the hyperbolic equation (\ref{nonlin-nondispersive}) and the circle to the Whitham equation (\ref{Whitham}).}
\label{fig:breaking_time_vor_Whitham}
\end{figure}

Since the KdV equation is not the appropriate model for describing wave evolution to breaking, \citet{Whitham} suggested as model the equation (\ref{Whitham}) (with $\Omega=0$). The Whitham equation and KdV equation have the same nonlinear term and different dispersive terms. The dispersive term of the Whitham equation corresponds to exact linear dispersion and consequently allows the introduction of small scales which are important in the breaking phenomenon. Constantin \& Escher(1998) have considered from a mathematical point of view breaking waves for Whitham-type equations without vorticity effect. They proved rigorously that a sufficiently asymmetric initial profile yields wave breaking. As emphasized by the latter authors, within the framework of weakly nonlinear waves we do not find a blow-up of the slope for symmetric initial profiles without vorticity as shown in figure \ref{fig:breaking_time_vor_Whitham}.
Due to imbalance in nonlinearity and dispersion for large values of $\Omega$, the breaking time is equal to that given by the hyperbolic equation (\ref{nonlin-nondispersive}). Note that for $\Omega$ larger than approximately $0.50$ breaking occurs. On the opposite, we found that a sufficiently asymmetric initial profile evolves to breaking without vorticity as shown in figure \ref{fig:breaking_time_vor_Whitham}. For that purpose we have considered the initial profile given by equation (\ref{wind_profile}) with $a=0.16$ and $\varphi \neq 0$. In that case the initial profile is asymmetric, typical of wind wave profiles with the leeward slope steeper than the backward slope, and breaking occurs for $\Omega$ larger than a negative threshold value close to zero ($ \approx -0.10$). Note that for amplitude $a<0.16$ and phase $\varphi=0$ we did not observe a blow-up of the slope of the leeward side of the crest. In addition to the asymmetry of the wave profile a condition on the nonlinearity of the initial profile is required to obtain wave breaking. Consequently, for an insufficient asymmetry of the initial condition an increase of its amplitude is required to observe breaking of the wave. Negative values of the vorticity stimulate the breaking phenomenon (opposing current) even though breaking occurred for weak positive values of the vorticity (advancing current). Generally, we can observe that there is no blow-up of the leeward slope when $\Omega < 0$. In that case, because the wave propagates on advancing current, we can expect that both the waveheight and wave slope will not increase. We have run numerical simulations for several negative values of $\Omega$ with symmetric and asymmetric initial wave profiles. The long-time evolution of the wave profiles has shown that the waveheight and leeward slope do not increase. The effect of the advancing current is to prevent the blow-up occurrence. Note that in the case of weak negative values of $\Omega$ the blow-up occurs. In that case, the intensity of advancing current is not sufficiently strong to prevent the onset of breaking wave.
\newline
In order to satisfy the criterion of weakly nonlinear waves we have reduced the value of the amplitude of the initial condition and increase the asymmetry of its profile. A forcing term $F$ is applied during a short period of time which consists of a sine progressive wave with the phase velocity $c$, in quadrature with the surface elevation $\eta(x,t)$.
\begin{eqnarray}
F=\epsilon \sin t \sin (x-ct), \qquad 0 \leq t \leq \delta t \nonumber \\
F=0, \qquad \qquad \qquad \hspace{1.9cm}  t > \delta t \nonumber
\label{forcing}
\end{eqnarray}
with $\epsilon=0.50$, $c=1$ and $\delta t =0.8$.
\newline
The initial condition is given by equation (\ref{wind_profile}) with $a=0.10$ and $\varphi=0$, and the forcing $F$ is applied during $\delta t$. In that case a blow-up of the slope is obtained. We can conclude that within the framework of weakly nonlinear waves a sufficient asymmetry of the initial profile yields to wave breaking as demonstrated by Constantin \& Escher (1998).
\newline
Figure \ref{fig:breaking_time_vor_Whitham} shows that the breaking time decreases as $\Omega$ increases.

\subsection{Within the framework of the generalised Whitham equation}

\begin{figure}
\vspace{0.2cm}
\center
\includegraphics[width=0.6\linewidth]{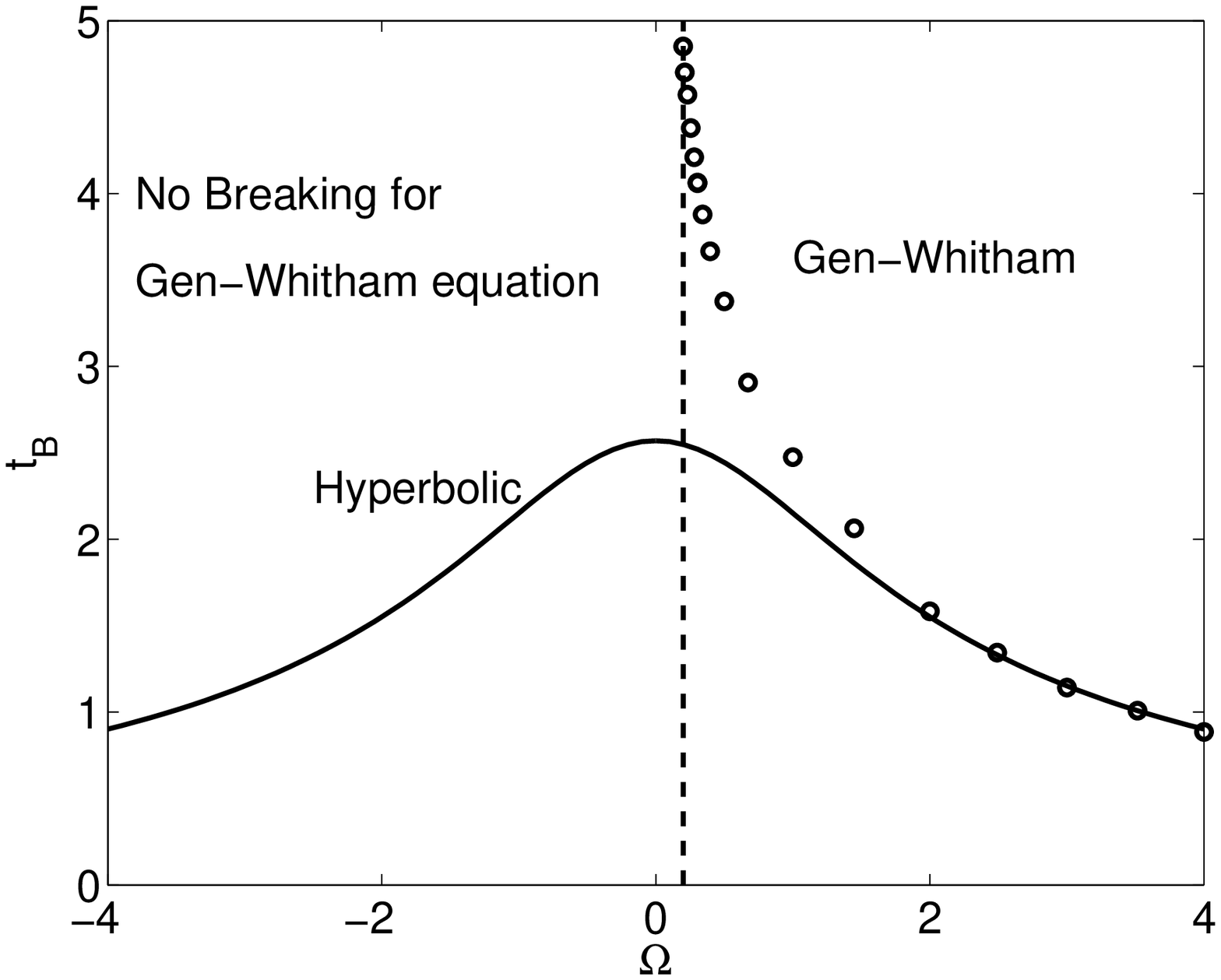}
\includegraphics[width=0.6\linewidth]{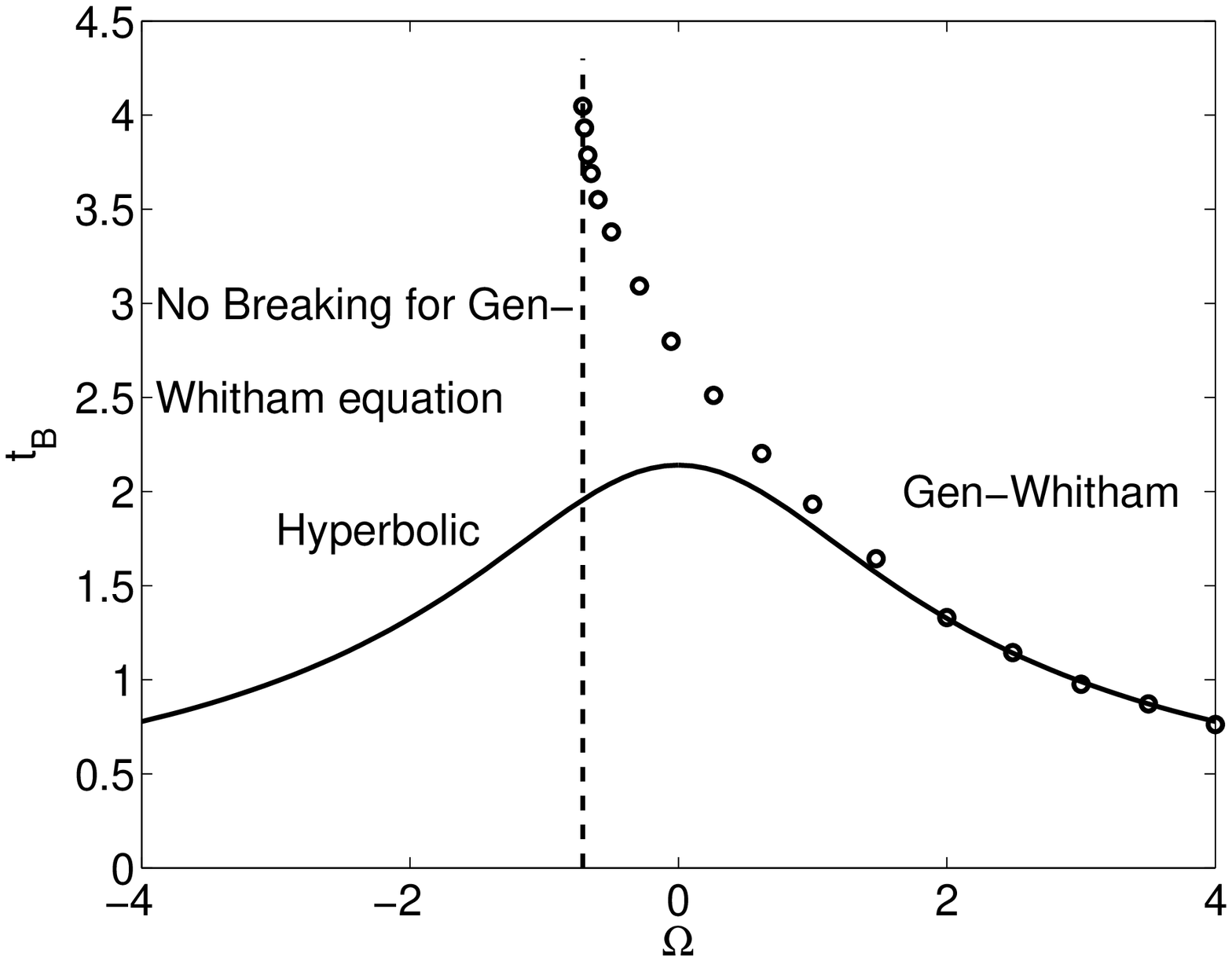}
\caption{Breaking time as a function of the vorticity within the framework of the generalised Whitham equation with a symmetric initial condition ($a=0.20, \, \varphi=0$) (top) and an asymmetric initial condition ($a=0.20, \, \varphi=3 \pi/2$) (bottom). The solid line corresponds to the hyperbolic case.}
\label{fig:breaking_time_Gen_Whitham}
\end{figure}

\begin{figure}
\vspace{0.2cm}
\center
\includegraphics[width=0.6\linewidth]{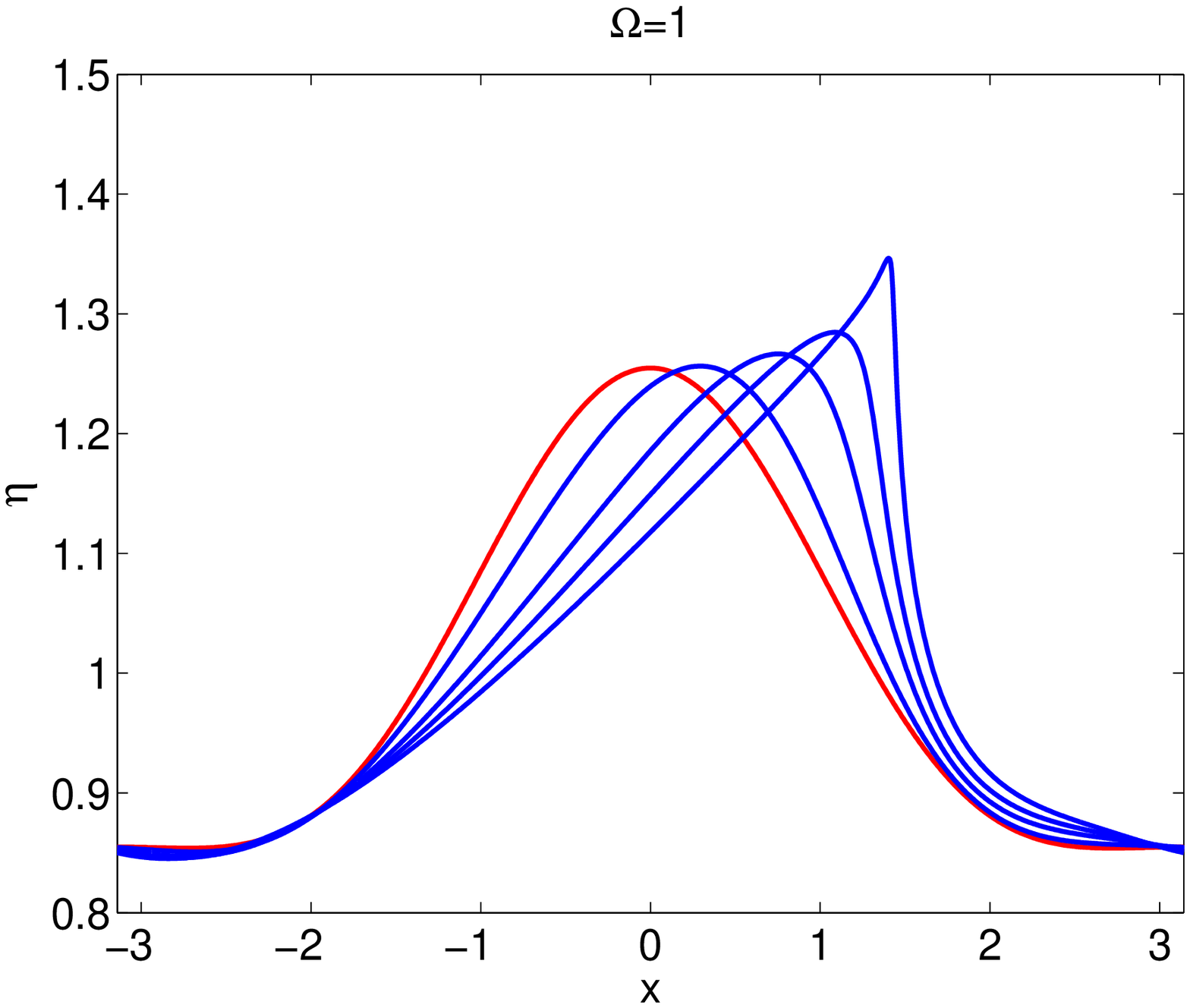}
\includegraphics[width=0.6\linewidth]{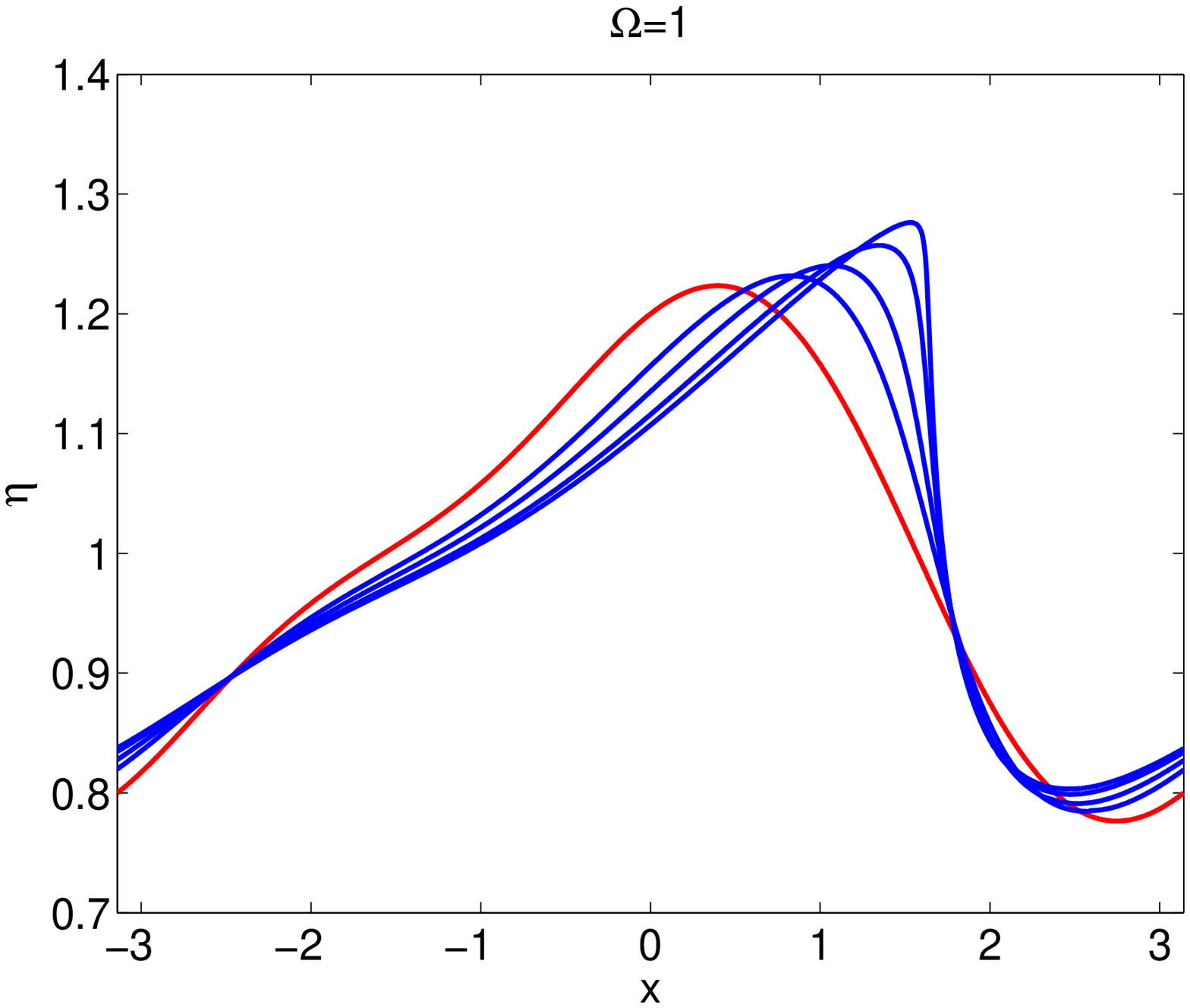}
\caption{Time evolution of the initial symmetric profile ($a=0.20, \, \varphi=0$) (top) and asymmetric initial profile ($a=0.20, \, \varphi=3 \pi/2$) (bottom) to breaking for $\Omega=1$.}
\label{fig:evolution_profile_Om1_Gen_Whitham}
\end{figure}

The Whitham equation governs the evolution of weakly nonlinear waves whereas the generalised Whitham equation describes the evolution of fully nonlinear waves. In figures \ref{fig:breaking_time_Gen_Whitham} is plotted the breaking time as a function of the vorticity for symmetric and asymmetric initial profiles given by equation (\ref{wind_profile}), respectively. For an initial amplitude $a=0.20$ we observe the blow-up of the slope for both symmetric and asymmetric initial profiles as soon as $\Omega$ becomes larger than a given threshold ($0.20$ and $-0.71$ for the symmetric and asymmetric initial conditions, respectively). Note that for this value of the amplitude the symmetric initial profile does not evolve to breaking in the absence of vorticity dispersive effects prevent the blow-up occurrence of the slope. The trend observed within the framework of the Whitham equation is amplified due to stronger nonlinearity. Like in the case of the Whitham equation, sufficiently powerful advancing current can eliminate the onset of breaking wave. 
\newline
Figure \ref{fig:breaking_time_Gen_Whitham} shows that the breaking time decreases as the current intensity $\Omega$ increases.
\newline
For $\Omega=1$, figures \ref{fig:evolution_profile_Om1_Gen_Whitham} show the profiles of the wave at different times before breaking for symmetric and asymmetric initial conditions, respectively. The amplitude and asymmetry of the profile increase as the wave evolves to breaking. We can observe the incipient formation of a jet at the crest of the wave.

\section{Conclusion}

Following \citet{Whitham} approach, we have derived from Riemann invariants, in the presence of constant vorticity, a new single partial differential equation governing the spatio-temporal evolution of the free surface elevation, then the velocity can be determined through an algebraic equation by using Riemann invariants. This system of partial differential equation and algebraic equation is fully equivalent to the St Venant equations for $1D$-propagation and allows the resolution of the evolution of hyperbolic waves of arbitrary height in the presence of vorticity. To take account of dispersion, we have introduced in the previous partial differential equation the exact linear dispersion of gravity waves on finite depth and in the presence of vorticity. From the latter equation we have derived the Whitham equation for weakly nonlinear water waves propagating in the presence of constant vorticity. Under the assumption of weakly nonlinear and weakly dispersive waves we have rediscovered in more straightforward manner the KdV equation with vorticity derived previously by \citet{Freeman}  and \citet{Choi}. We found that solitary wave and cnoidal wave profiles are broader for positive vorticity whereas they are narrower for negative vorticity. Furthermore, positive vorticity increases their phase velocity whereas it is the opposite in the presence of negative vorticity. The Whitham equation with non zero vorticity has been tackled, too.
\newline
To follow the evolution of non breaking undular bores, we have implemented a pseudo-spectral numerical method. The validity of the numerical method has been checked against the exact analytical solution of the St-Venant equations for the dam-break problem in the absence of current and vorticity. A second validation has been carried out by using Favre's experiments on undular bores with frictional effect at the bottom and without vorticity.
\newline
A particular attention has been paid to non breaking undular bores in the presence of vorticity, too.  We have shown that negative vorticity increases the height of the leading wave whereas it decreases in the presence of positive vorticity. The wavelength of the following waves is also modified by vortical effects, it is shortened in the presence of negative vorticity and stretched for positive vorticity. 
Within the framework of deep water, \citet{Touboul} using a Boundary Integral Element Method found similar results.
\newline
An investigation on the breaking time of dispersive waves in the presence of constant vorticity has been carried out within the framework of the Whitham equation and generalised Whitham equation, respectively. The numerical method to capture the blow-up of the slope corresponding to the onset of the breaking has been checked against the analytical solution of the St-Venant equation in 1D. Numerical and exact results are in excellent agreement. We found that the breaking time of hyperbolic waves decreases as the vorticity magnitude increases whatever its sign. The breaking time of dispersive waves decreases as the shear intensity, $\Omega$, increases. However, the cases we have considered have shown that dispersive waves propagating in the presence of a strong positive vorticity ($\Omega<0$) do not evolve to breaking.
\vspace{1cm}
\newline
{\bf Appendix A. Periodic travelling-wave solutions of the Whitham equation}
\vspace{0.3cm}
\newline
The Whitham equation (\ref{Whitham}) is  rewritten as
\[
\eta_{t}+c_{1}(\Omega)\eta\eta_{x}+K*\eta_{x}=0,\quad c_{1}(\Omega)=\frac{3gh+h^{2}\Omega^{2}}{h\sqrt{gh(4gh+h^{2}\Omega^{2})}}.
\]
 Travelling-wave solutions of the form $\eta(x,t)=\phi(x-ct)$, for
a given phase velocity $c$, are sought. They are solutions of the stationary
equation:
\[
-c\phi+c_{1}(\Omega)\frac{\phi^{2}}{2}+K*\phi=0.
\]
 For $\Omega=0$, \citet{Ehrnstrom} proved the
existence of a branch of $2\pi$-periodic traveling-wave solutions
for the Whitham equation. We will show, numerically, the existence
of $2\pi$-periodic branches of solutions for $\Omega\ne0.$ Let the
approximate solution be: 
\[
\phi_{N}(X)=\sum_{n=0}^{N}a_{n}\cos(nX),\quad X\in[0,2\pi].
\]
 The residual is then: 
\[
R_{N}(a_{0},a_{1},...,a_{N})=-V\phi_{N}+c_{1}(\Omega)\frac{\phi_{N}^{2}}{2}+K*\phi_{N}.
\]
 We use a pseudo-spectral method. The nonlinear term is computed in
the physical space ($[0,2\pi]$ is discretized using a regular grid).
The convolution is evaluated in the spectral space using Fast Fourier
Transforms. To make the residual minimal, we use a Galerkin method:
\[
\langle R_{N},\cos(nX)\rangle=\int_{0}^{2\pi}R_{N}\cos(nX)dX=0,\quad n=0,N,
\]
 and this gives $N+1$ nonlinear equations for the unknowns $a_{n}$.
These equations are solved using the fsolve routine from matlab. The
initial guess is a cosine wave with a small amplitude and a phase velocity
$c$ estimated using the linear dispersion relation. We used $N=20$,
and the computations are stopped when the residual norm is $O(10^{-14})$.
Then, we employ a continuation method to obtain solutions for different
values of $c$. For $\Omega=0$ we checked that our results are in
agreement with those of Sanford et al \citet{Sanford}. The results
of computations for different values of $\Omega$ are given in section 2.
\vspace{0.3cm}
\newline
{\bf Appendix B. Detection of blow-up occurrence in finite-time}
\newline
The blow-up after a finite time is detected using the method of the analyticity strip presented in \citet{Sulem83}. The essence of the method is the following.
When $\eta(x,t)$ is an analytic function, its Fourier coefficients (with respect to $x$) decay faster than any power of $1/k$ ($k$ is the wavenumber) in the limit $k\rightarrow\infty$. When $\eta$ is singular, its Fourier coefficients decay algebraically with $1/k$. Hence, to detect the time of the appearence of the singularity we assume that the Fourier coefficients of the solution $\eta(x,t)$ behave as:
$$\hat\eta_k(t)=C(t)k^{-\alpha(t)}e^{-\delta(t)k}.$$ 

The adjustable coefficients $C$, $\alpha$, $\delta$ are calculated using a least sqaure method. The coeffecient $\delta$ is known as the analyticity strip width. Loss of regularity correponds to the vanishing of $\delta(t)$ and the corresponding time gives the time of breaking $t_B$. This method is validated against the time of breaking when all the equations studied here are hyperbolic (vanishing dispesion). In those cases, the expression of the time of breaking could be obtained analytically as a function of $\Omega$. An example of this validation is given in figure \ref{fig:breaking_time_hyperbolic}.
 
%
%
%
%

\end{document}